\documentclass[fleqn,12pt]{wlscirep}
\usepackage[utf8]{inputenc}
\usepackage[T1]{fontenc}
\usepackage{bm}
\usepackage{tabularx} 
\usepackage{color,comment}
\usepackage{subfigure}
\usepackage{framed}
\usepackage{multirow}

\title{Tipping points in complex ecological systems}

\author[1]{Alan Hastings}
\author[2]{Sergei Petrovskii}
\author[2]{Valerio Lucarini}
\author[2,*]{Andrew Morozov}
\affil[1]{University of California, Department of Environmental Science and Policy, Davis, USA and Santa Fe Institute, Santa Fe, USA}
\affil[2]{University of Leicester, School of Computing and Mathematical Sciences, Leicester, UK}

\affil[*]{e-mail: am379@le.ac.uk}

\begin{abstract}
Tipping points are one of the hot topics \textcolor{black}{in the modern physics} of complex systems. But what is a tipping point? \textcolor{black}{One generic definition describes} it as ``a state of the system where a small change in its parameters can lead to a significant change in its properties''. Additional ingredients that often enter the definition of tipping point are the abruptness of the resulting change and its irreversibility, i.e. it is impossible to recover the initial state if one reverses the protocol of change of the parameters. However, there exists a number of different mathematical structures that can show this behavior, \textcolor{black}{the one that was originally used (nowadays often referred to as the bifurcation induced tipping) is just one of many}. 
Different preconditions and/or different level of \textcolor{black}{detail} included into the model, reflecting also different environmental forcing, can lead to a variety of tipping mechanisms. 
Furthermore, in a spatially extended system and/or a system with multiple scales, different parts can react to a change in environmental conditions differently or at a different time, interacting with each other to create a tipping cascade. 
In this paper, {\color{black}using ecosystems as a paradigm} of complex nonlinear open systems, we provide a critical overview of the progress made in tipping point science over the last 15 years. We highlight the main findings, identify gaps in our knowledge, and outline a roadmap for further progress.

\end{abstract}
\begin{document}

\flushbottom
\maketitle

\thispagestyle{empty}

\noindent \textbf{Key points:} 
\begin{itemize}
    \item Understanding, anticipating, and predicting tipping points is one of the key research challenge for the study of ecological systems 
    \item The phenomenology of \textcolor{black}{tipping points} is complex and requires a vast portfolio of mathematical techniques and ideas
    \item The convergence of data-driven methods and theory-informed strategies is leading to encouraging progress
    \item Tipping is increasingly seen as a complex phenomenon where the emergence of spatial patterns can play a prominent role
    \item Evidence for tipping in both empirical observations and in models is found in a number of important ecological systems.
\end{itemize}

\noindent \textbf{Website summary:} Tipping points are sudden transitions in complex {\color{black}nonlinear} systems that present unique challenges for theory and empirical work.  We emphasize the role of timescales, spatial complexity and complexity of {\color{black}ecological} interactions.  We outline recent theoretical progress that apply to these systems and provide empirical examples.

\section*{Introduction}

\begin{figure}[!b]
\centering
\subfigure[]{\includegraphics[scale=0.25,angle=0]{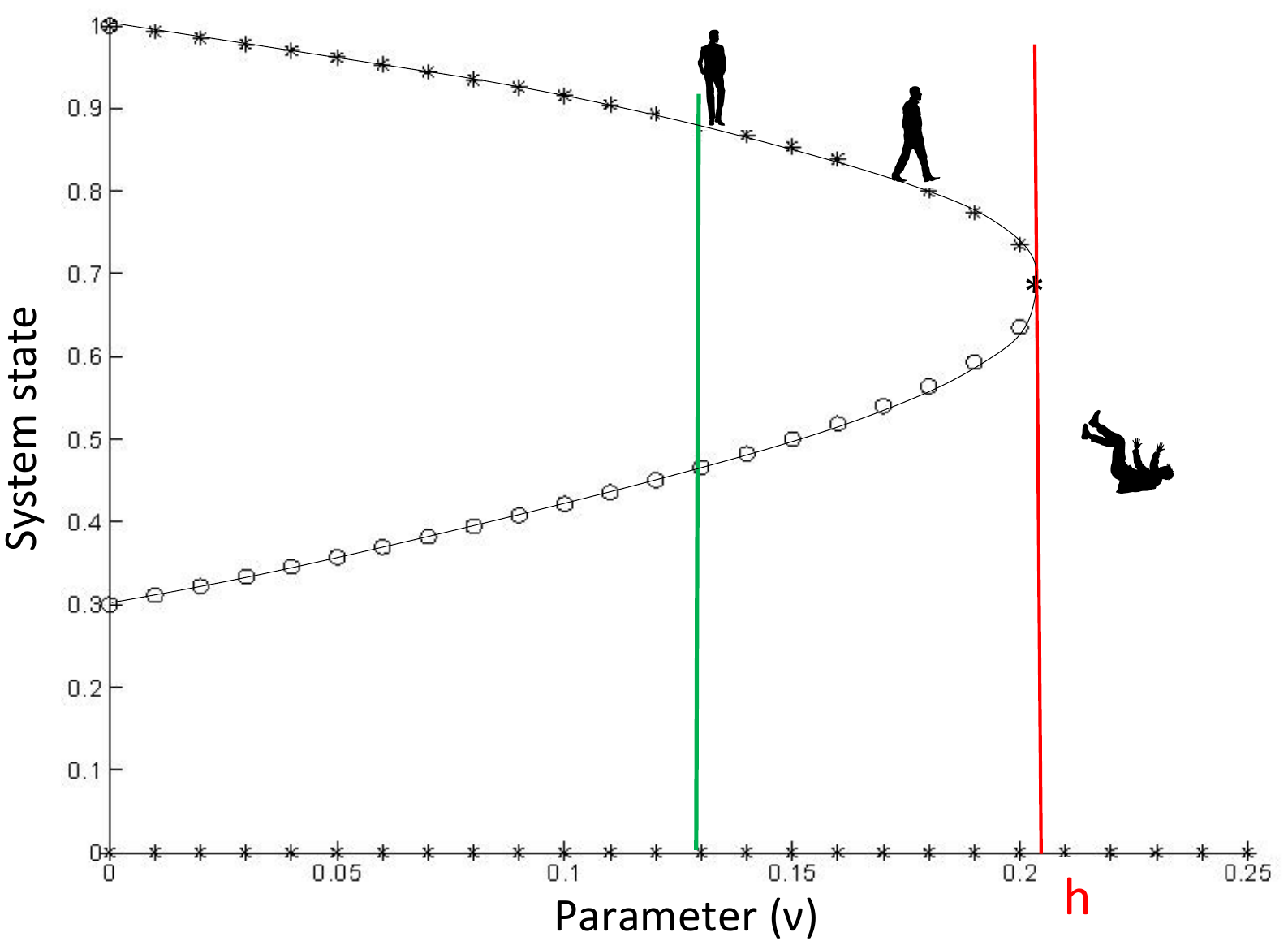}}\hspace{5mm}
\subfigure[]{\includegraphics[scale=0.25,angle=0]{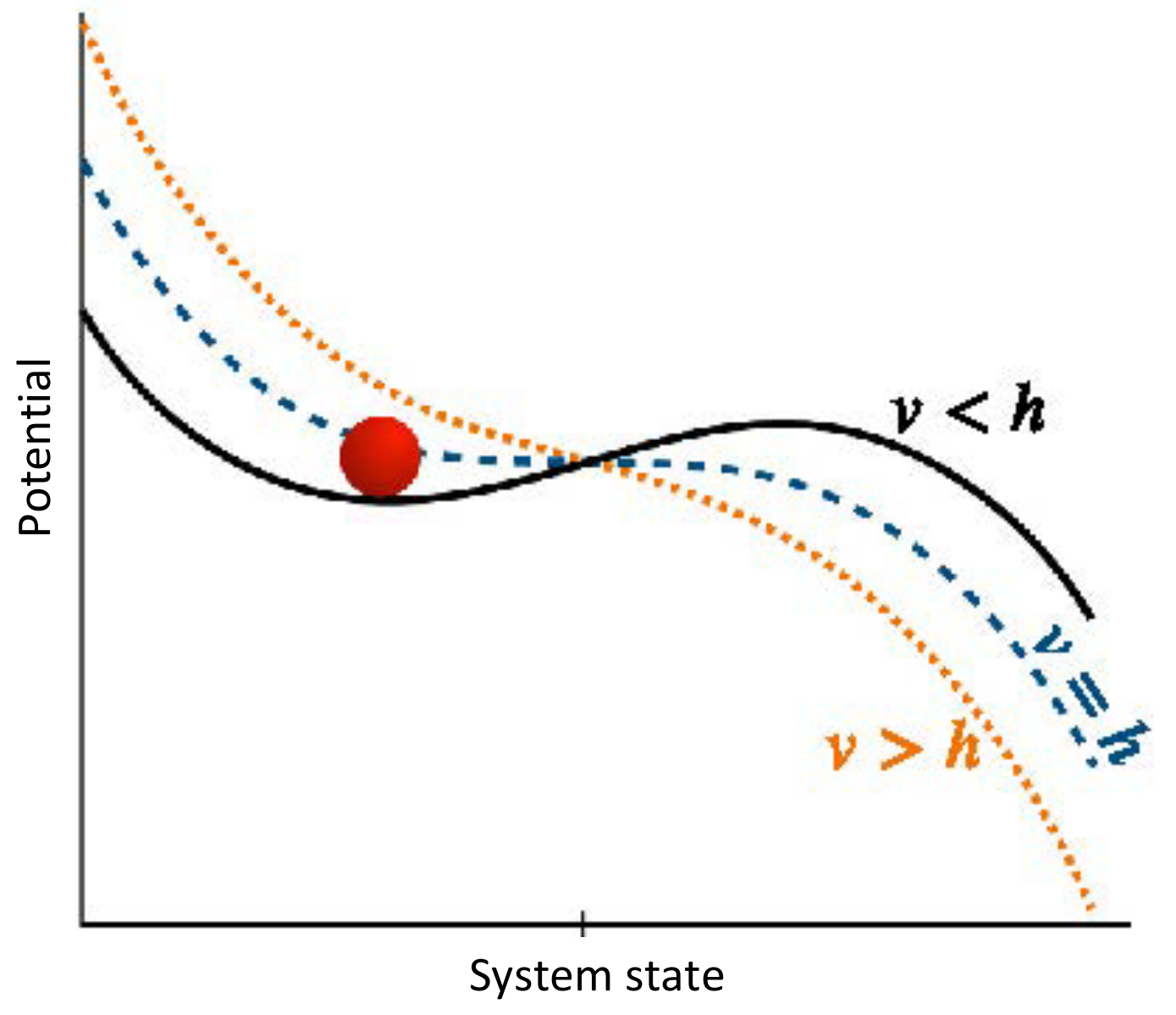}}
\caption{\small Tipping point in a prototypical one-dimensional system of the form $\dot{x}=f(x,\nu)=-\mathrm{d}V(x,\nu)/\mathrm{d}x$ with a few steady states. (a) A stable steady state of the system (shown by asterisks / the upper branch of the curve) can merge with an unstable state (shown by open circles / lower branch of the curve) and disappear when a system's parameter $\nu$ crosses its critical value, say $\nu=h$ (shown by vertical red line). The system is therefore in a `safe' state for any $\nu<h$ (as is visualized by the green line) but will experience a `free fall' transition to another state for $\nu>h$. (b) The disappearance of the steady state in terms of  $V(x,\nu)$: a sufficiently (overcritical) increase in the bifurcation parameter $\nu$ will eliminate the local minimum, {\color{black}so that the system transitions to an alternative state (not shown here)}. Figure adapted from\cite{abbott2024}.}
 \label{Figure1}
\end{figure}

Over the last two decades, progress in studies on environmental systems (including climate and ecosystems across multiple scales), supported by the progress in understanding general properties of nonlinear dynamics and complex systems, has opened a new perspective on critical transitions by placing them into the context of  dynamical systems theory. The system `tips' or crosses a tipping point when its reference stable steady state either  disappears\cite{scheffer2001,ashwin2012tipping} {\color{black}or losses it stability via a global bifurcation\cite{Thompson2011predicting}}. The most studied case is associated  with the occurrence of a saddle-node bifurcation when a relevant parameter passes its bifurcation value, see Fig.~\ref{Figure1}a. Once it happens, the system transitions to another steady state (or a stable dynamical regime, e.g.~self-sustained oscillations); in the ecological context, such regime shift is often associated with biodiversity loss and ecosystem degradation. 
If one considers simple one-dimensional models of the form $\dot{x}=f(x,\nu)$, where $\nu$ is a parameter, this can also be understood in terms of a `stability landscape' due to potential function. Since we can write $\dot{x}=f(x,\nu)=-\mathrm{d}V(x,\nu)/\mathrm{d}x$, the bifurcation described above  occurs when a change in the parameter $\nu$ leads to eliminating the local minimum of $V$ associated with the reference state (Fig.~\ref{Figure1}b).
{\color{black}For high-dimensional, non-gradient systems (to which ecosystems belong) potential function usually does not exist but tipping in response to a parameter change can be understood in a similar way using, instead, the notion of quasi-potential \cite{zhou2012,Cameron2012}. } 

The above tipping scenario, now commonly known as the bifurcation tipping (B-tipping) includes some generic features of critical transitions such as the existence of a critical threshold, but it has much more to offer. B-tipping eliminates the present state, so the system has to transition to an alternative one. This is similar to a `usual' critical transition in a physical system.
However, one important difference (and, as we will show it below, not the only one) emerges from the fact that, in a certain range of environmental conditions (i.e.~for the same value of a controlling parameter, e.g.~temperature or precipitation), {\color{black}even in the nonspatial (spatially uniform) case}
there may exist {\it multiple stable states} of the system. One of the best known examples is given by the savanna-forest bistability\cite{staver2011}, more examples will be given in the ``Paradigmatic case studies'' section. 

{\color{black}The B-tipping scenario described above and visualized in Fig.~\ref{Figure1} is a baseline one which, strictly speaking, can only occur in autonomous systems.} 
When the above analysis is applied to an ecological (environmental) system, a change in parameter values is usually associated with a change in the environmental conditions. 
{\color{black}It means that parameter $\nu$ becomes a function of time, $\nu=\nu(t)$. In this case, the system becomes nonautonomous and 
scenarios of tipping become diverse and considerably more complicated. However, in case the rate of change in the parameter value is slow compared to the intrinsic timescales of the system, the change in the system's properties can still be analyzed using a sequence of `frozen' autonomous systems similar to the above.} 

Whilst most visual representations of tipping are by necessity one-dimensional, the scenario depicted in Fig. \ref{Figure1} can be generalised also for  high-dimensional systems. If the system is described by a gradient dynamics, $\dot{\mathbf{x}}=\mathbf{f}(\mathbf{x},\nu)=-\mathbf{\nabla} V(\mathbf{x},\nu)$, $\mathbf{x}\in\mathbb{R}^N$ where $N>1$, B-tipping occurs when, for a critical value of the parameter $\nu=h$, the local minimum merges with a saddle. Importantly, the scenario extends to the much more general and relevant case of non-gradient multistable systems, including those possessing competing states featuring chaotic dynamics. {\color{black}In this case, B-tipping occurs as a basin crisis \cite{Grebogi1983} associated with the reference attractor touching an edge state \cite{Skufca2006,Bodai2020} living in the basin boundary \cite{LucariniBodai2017}. The edge state is a high-dimensional dynamical saddle embedded in the basin boundary, featuring one transversal unstable direction. As a result, whilst being an invariant set, the edge state is only a relative attractor: it attracts trajectories initialised on the basin boundary. However trajectories initialised in a neighborhood of the edge state escape such a neighborhood according to an exponential distribution of life times \cite{Lai2011}.} If the dynamics on the edge state is chaotic, one might obtain the puzzling result that the basin boundary separating the competing stable states is fractal, having a codimension strictly smaller than one, and possibly very close to zero \cite{LucariniBodai2017,Bodai2020}.
At the tipping point, the merging of the edge state and the reference attractor leads to the emergence of a ghost state \cite{Strogatz1989,Deco2012,Madeiros2018}.  {\color{black}Note that the concept of B-tipping also applies to some non-autonomous systems, in particular, those including periodic forcing to describe daily and seasonal variations of parameters \cite{wollrab2021climate}. Bifurcation theory for non-autonomous ODEs with periodic forcing can be applied in this case. Alternatively, one can sample the continuous flow at discrete time intervals equal to the forcing period and consider B-tipping of the corresponding autonomous time-discrete stroboscopic map.}

It is possible to a make further step and introduce in this case a notion of quasi-potential \cite{Graham1991,zhou2012,LucariniBodai2020}, which  generalises the potential $V$ discussed above to the case of non-gradient systems. Practical construction of this scalar function, which summarizes the stability landscape of the system, requires adding stochastic forcing to the system, which will be dealt with further in the paper. 

Questions may arise here - what is so special about ecosystems? Can they bring a new challenge to the science of tipping points that might not be present in other natural systems? As it happens, ecosystems indeed possess many dynamic and structural features that distinguish them from other natural systems, as is briefly outlined in Box 1.
{\color{black}Below we will see how many of these differences play out in canonical examples of tipping in ecosystems including coral reefs \cite{Mumby_etal_2007,Hughes2003,Hughes2013} and lakes \cite{scheffer2001,carpenter2011early}}.
Combinations of the above factors and an interplay between them make ecosystems highly complex systems, arguably more complex than any other natural system. This complexity has a variety of implications, in particular it opens a possibility of tipping mechanisms different from classical B-tipping. 
Some of them are briefly outlined in the next section. 

\begin{figure*}[!ht]
\begin{framed}
\label{box1}

{\footnotesize

{\bf Box 1. Uniqueness of ecosystems as complex systems}

\vspace*{1mm}

{\it Foodweb complexity.} Prey-predator-type interactions along with interspecific competition bring different species together into a trophic network, i.e. a foodweb\cite{polis1996}. A typical foodweb consists of hundreds or even thousands of nodes (species) that are all unique and the network can have a complicated structure\cite{dunne2002}.

\vspace*{1mm}

{\it Broadness of relevant spatial scales.} Due to the difference in the size of individual organisms (of species that are coupled together through trophic interactions, hence being parts of the same foodweb) and/or their typical dispersal distances, relevant spatial scales\cite{levin1992,zelnik2024} can span across ten orders of magnitude, ranging from $10^{-4}$ meter (e.g. the size of microplankton in marine ecosystems) to  $10^6$ meter (dispersal distance of some whales, birds, fish, etc.)

\vspace*{1mm}

{\it Complexity of temporal dynamics.} Complex intrinsic (self-organised) dynamics of interacting species, which includes periodic or quasi-periodic oscillations\cite{myers2018}, deterministic chaos\cite{hastings1993}, intermittency\cite{Monti2024}, etc. is further complicated by various external forcing\cite{Amritkar2006,Chekroun2006,Roques2007} that can be deterministic and periodic (e.g. through diurnal or seasonal changes) or stochastic (oscillations in weather and climate, fluctuations in the solar activity, random anthropogenic interventions, etc.)

\vspace*{1mm}

{\it Multiplicity of temporal scales.} Complexity of temporal dynamics is further complicated by the existence of multiple timescales\cite{kuehn2015}. In ecosystems, the latter can have various origin\cite{Hastings2010}; in the simplest case, it is provided by the differences in baseline timescales such as species average life-span and the time between consequent generations. In an ecological community, these can span across a few orders of magnitude, ranging from hours (e.g. in seasonal phytoplankton reproduction) to decades (for some avian and reptile species). 

\vspace*{1mm}

{\it Timescale of response.} As a marked contrast to the analysis of phase transitions in physics which basically are changes in asymptotic behavior \cite{sole2011phase}, in ecology the dynamics in the neighborhood of a tipping point can play out slowly \cite{Hastings2010,Hughes2013}.  Thus, a system which essentially has already tipped and is on a path to end up in a different state if conditions did not change could be rescued by relatively quick action {\color{black} as has been emphasized for coral systems \cite{Hughes2013}.}{\color{black} Time scales of response show up in other ways. The importance of global change for ecosystems which leads to nonautonomous systems means that rate induced tipping  is important\cite{Kiers2020,abbott2024}.}

\vspace*{1mm}

{\it Spatial complexity.} The environment is heterogeneous on multiple scales. In terms of the species habitat or ecosystem type, this heterogeneity often leads to fragmentation when intrinsically continuous physical space appears to be split to a number of coupled patches, usually referred to as metapopulation\cite{hanski1998}. In turn, it leads to new phenomena, for instance a possibility of synchronization between population oscillations at different locations\cite{blasius1999,noble2015}. In a homogeneous environment, nonlinear interspecific interactions can result in self-organized pattern formation and spatiotemporal chaos\cite{Medvinsky2002,malchow2008}. 

\vspace*{1mm}

{\it Active species feedback on their environment.} This can occur on different spatial scales, from local (``ecosystem engineers''\cite{Hastings2007}, e.g.~beavers) to global (changes in the global energy balance through changes in surface albedo\cite{Charlson1987,tian2014}). In a slightly different context, the ability of species to amend their environment according to their needs is known as the Gaia hypothesis\cite{Lovelock2000}.  

\vspace*{1mm}

{\color{black}
{\it Physiological response and phenotypic plasticity.} Contrary to abiotic components, biological species are capable to change their traits and functions in response to changing environmental conditions\cite{chevin2010}. Such phenotypic plasticity is generally believed to increase species fitness and hence can be expected to contribute to ecosystem resilience\cite{Dakos2019}. However, consequences of phenotypic evolution can be counter-intuitive, sometimes leading to a tipping point rather than mitigating it\cite{Fellows2026bioRrxiv}. 
Also, the response can be taxon-specific and for some animal taxa may result in a strong Allee effect\cite{Courchamp1999}, which in turn may lead to species extinction\cite{Miller2004}.}

\vspace*{1mm}

{\color{black}
{\it Data challenges.} Monitoring of ecological systems can be difficult.  Many methods proposed, for example, for early warning signs of impending tipping depend on data of a length and quality that are not available in ecological systems.}

\vspace*{1mm}

The above list is not exhaustive and can be further extended\cite{turchin2003,sole2022}. 
 }

\end{framed}
\end{figure*}

\section*{Types and scenarios of tipping}

Research in the last decade has already considerably broadened the original understanding of tipping mechanisms. 
B-tipping, as described above, is a paradigmatic scenario of abrupt changes in complex ecological systems \cite{scheffer2001,rietkerk2004}.  
However, it is not the only possible one. Indeed, the B-tipping mechanism  apparently misses many factors that are endemic in ecosystems, for example the effect of stochastic forces and perturbation, and the existence of different timescales. 
A number of alternative dynamical mechanisms have been identified that can lead to ecosystems tipping (see Table \ref{table1}). As we discuss below, they differ significantly in terms of the preconditions required for them to occur and hence also differ in terms of their applicability and generality. Of course, Nature does not follow theorems nor obey asymptotic limits, so that any real-life example of tipping will in general have features belonging to more than one of the idealised dynamical mechanisms. 

\begin{figure}[!h]
\centering
\subfigure[]{\includegraphics[scale=1.1,angle=0]{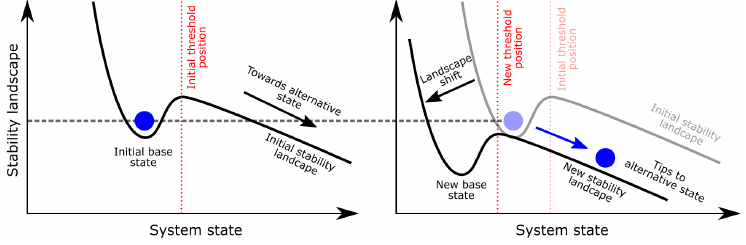}}
\subfigure[]{\includegraphics[width=0.41\textwidth,height=0.22\textwidth,angle=0]{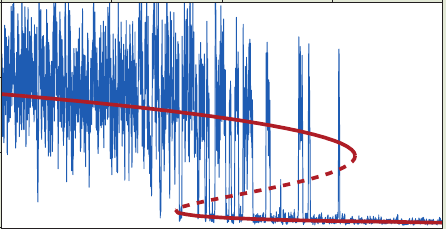}}\hspace{7mm}
\subfigure[]{\includegraphics[scale=0.89,angle=0]{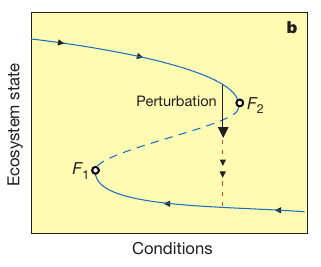}}
\subfigure[]{\includegraphics[scale=0.7,angle=0]{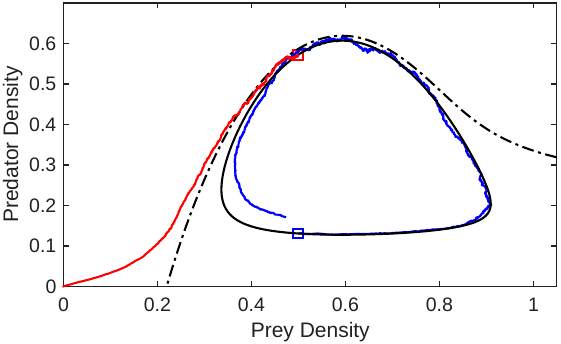}}\hspace{7mm}
\subfigure[]{\includegraphics[width=0.42\textwidth,height=0.25\textwidth,angle=0]{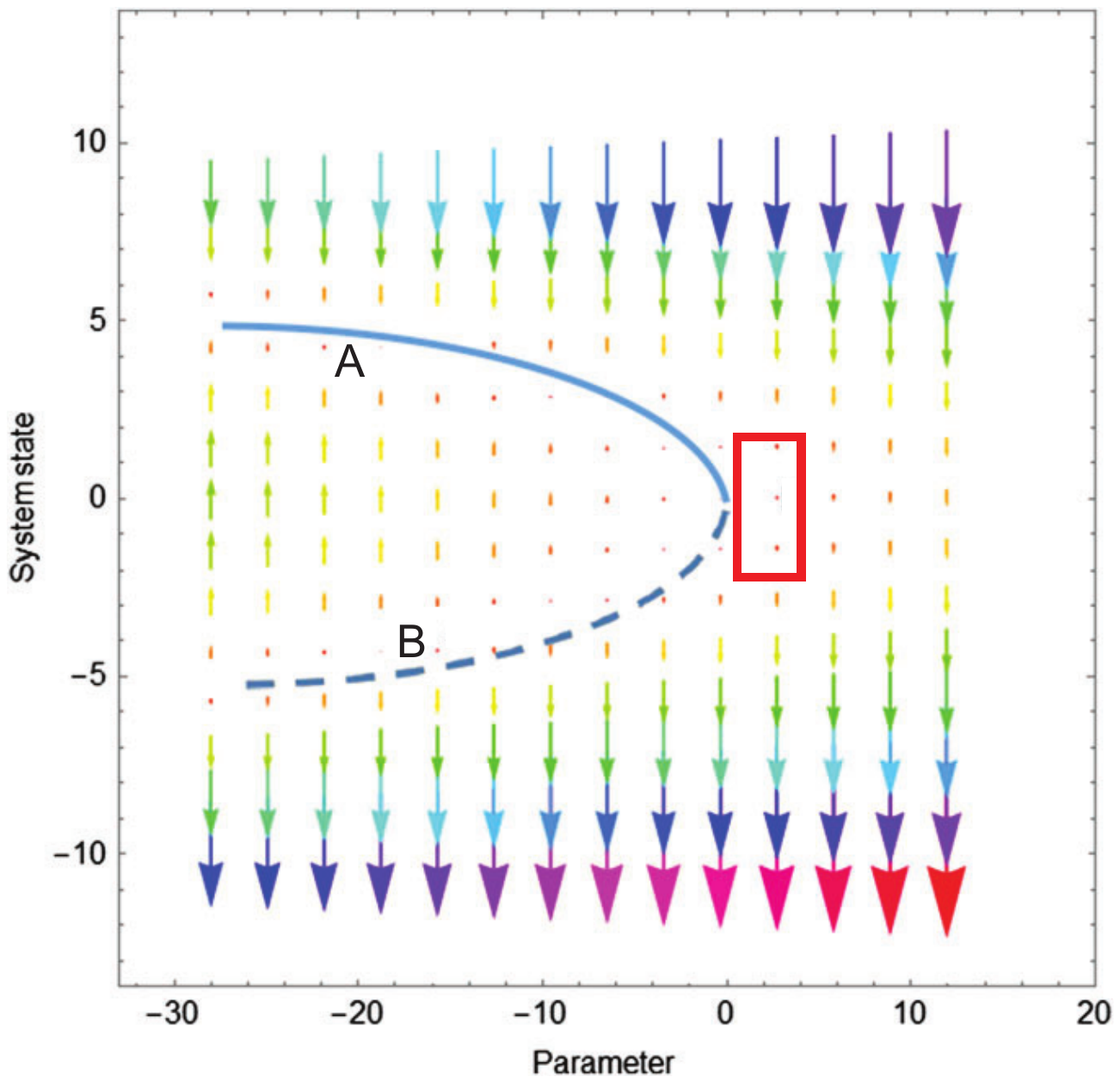}}
\subfigure[]{\includegraphics[width=0.37\textwidth,height=0.27\textwidth,angle=0]{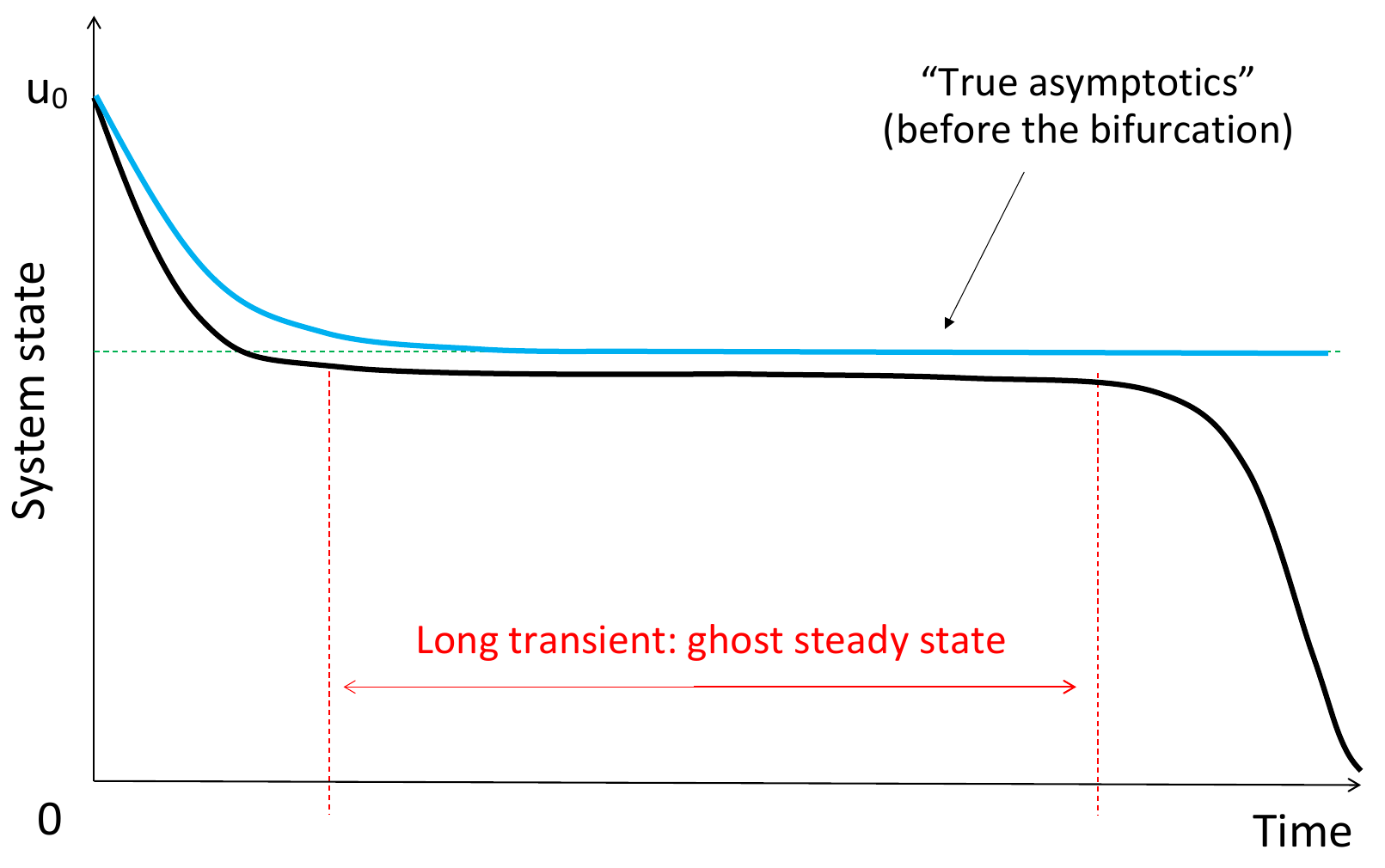}}\hspace{7mm}
\subfigure[]{\includegraphics[width=0.38\textwidth,height=0.27\textwidth,angle=0]{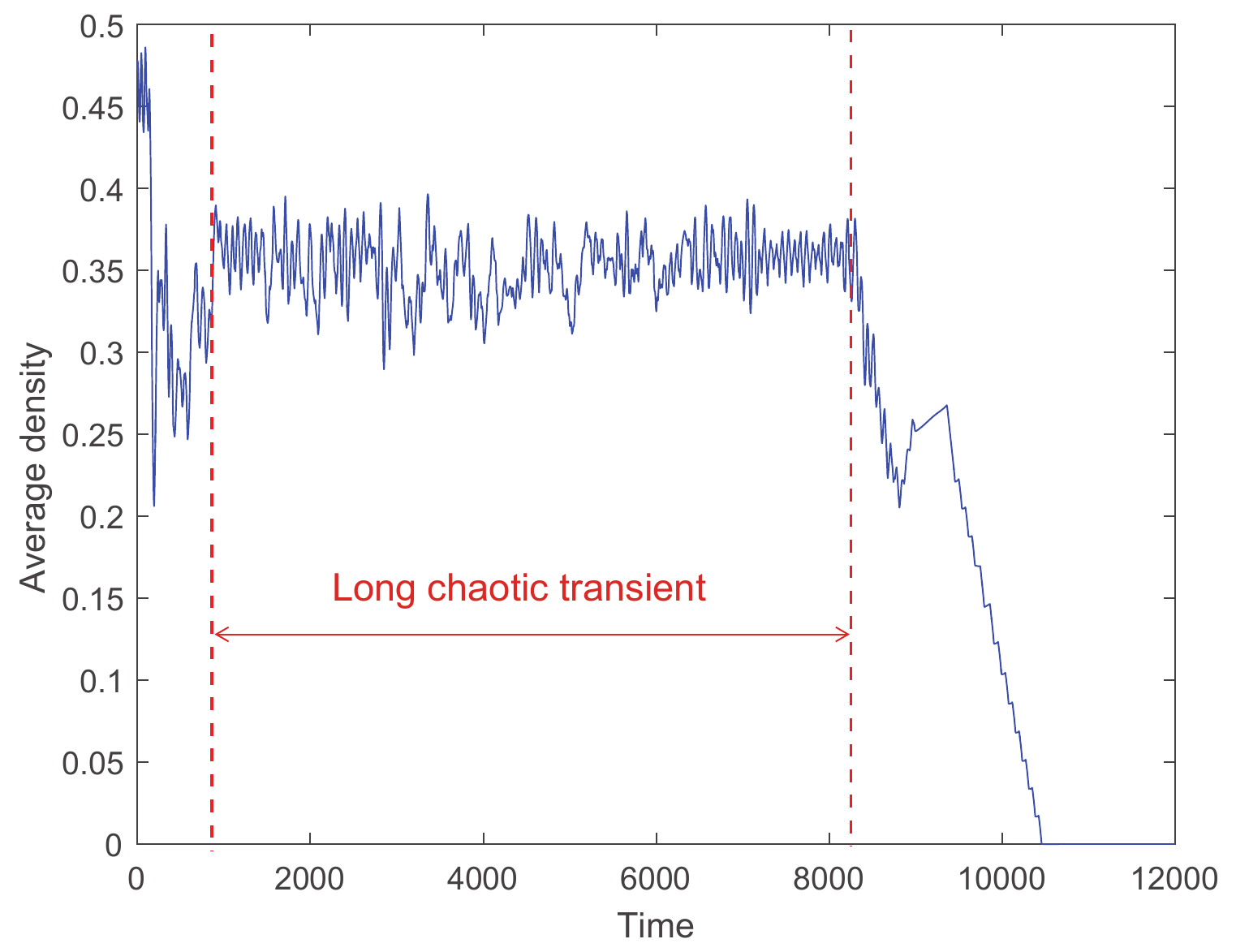}}
\caption{\small For the caption, see next page.}
 \label{Figure2}
\end{figure}

The B-tipping mechanism is based on several implicit assumptions. One such assumption is that the change in the controlling (bifurcation) parameter is infinitely slow. From a mathematical viewpoint, this avoids a specification of the explicit dependence of the bifurcation parameter on time by considering instead an ensemble of autonomous systems for different parameter values (up to and beyond its bifurcation point). That leaves entirely open the questions as to what can be the tipping scenario in a case where the change cannot be regarded as slow, e.g.~in the opposite extreme case of a fast parameter change. Note that the latter is highly relevant, because the rate of  ongoing climate change is believed to be too fast compared to typical timescales of ecosystem resilience. 
Consider the case where the value of a system parameter - say, $\nu$ – changes from $\nu_0$ to $\nu_1=\nu_0+\Delta\nu$ over a certain time interval $\Delta{t}$, i.e.~with the average rate $R=|\Delta\nu/\Delta{t}|$. It has been observed in several studies\cite{petrovskii2004,ashwin2012tipping,ritchie2023rate,osullivan2023,abbott2024,vanselow2024} that an ecological system can respond differently to perturbations of different rates: it may be resilient to a slow change, so that its dynamics and properties remain largely unchanged, but would experience a regime shift in the case of a sufficiently fast change. From the dynamical systems viewpoint, this mechanism (commonly referred to as rate-induced tipping or R-tipping\cite{ashwin2012tipping,Kiers2020}) can be thought of as a fast shift of the whole potential landscape while the position of the system in the landscape remains approximately the same (see Figure \ref{Figure2}a). As a result, the system may appear outside of its previous stability basin and hence will inevitably transition to an alternative stable steady state or regime (potentially resulting in drastic changes in ecosystem's properties, e.g.~species extinctions). {\color{black} Mathematically, this is described by the basin instability phenomenon \cite{ritchie2023rate}, where an instantaneous (frozen) attractor at time $t_1$ becomes located outside the basin of attraction of the transformed state at some later time $t_2>t_1$. In this case, the theory predicts the rate-induced tipping for a supercritical rate of parameter change \cite{ritchie2023rate,feudel2023rate}.}

\begin{figure}[h]
 \begin{framed}
{\small
{\bf Caption to Figure \ref{Figure2}.} Different tipping mechanisms. (a) Rate-induced tipping (R-tipping) occurs when a change in the environmental conditions occurs much faster than the system relaxation time. The system state remains approximately the same as before the change, however the whole quasi-potential landscape moves away~\cite{ritchie2023rate,abbott2024}. Figure adapted from~\cite{ritchie2023rate}. (b) Noise-induced tipping (N-tipping) occurs when a small stochastic perturbation (noise) occasionally push the system to an alternative steady state; the probability of such transition is the larger the closer is the system to the saddle-node bifurcation. Figure adapted from~\cite{scheffer2012}. (c) Shock-induced tipping (S-tipping) occurs when a single large perturbation can be enough to push the system out of its previous basin of attraction\cite{scheffer2001} (and hence to an alternative stable state or regime). Figure adapted from~\cite{scheffer2001}. 
(d) Phase-dependent tipping (P-tipping). In case the underlying system dynamics is periodic, quasi-periodic or chaotic, system's sensitivity to perturbations may depend on its `phase', i.e.~its position on the trajectory. The capacity of the same noise to lead (cf.~the red curve) or not lead (blue curve) to system's tipping then may depend on the timing of the perturbation.  
(e-g) Tipping due to the effect of long transients\cite{hastings2018transient}. (e) In the overcritical parameter range (shown by the red rectangle), there is no steady state but the dynamics is very slow. Figure adapted from~\cite{Hastings2016timescales}. (f) The slow dynamics may give an impression of a `ghost steady state'. (g) This holds as well for more complicated dynamics, in particular in the form of a `chaotic ghost'\cite{morozov2020,petrovskii2017,Mehling2024}. Figure adapted from~\cite{petrovskii2017}. }
 \end{framed}
\end{figure}

{\color{black}
We mention here that details of the R-tipping scenario can depend on the type of the attractor at which the dynamical system is before and/or after the tipping point. 
For instance, in case the system transitions from a steady state to a steady state, there is a well defined critical rate (say, $R_{cr}$) so that for any $R>R_{cr}$ the system leaves the previous attractor's basin (and hence a regime shift occurs). However, in case the system state before or after tipping is more complicated than a steady state, e.g.~periodic or chaotic, the rate dependence becomes nonsharp and can even become nonmonotonic. In order to quantify this change in the system's response - from a single critical threshold to a distribution of critical values - a concept of partial tipping was introduced\cite{Alkhayuon2018}. Interestingly, partial tipping can also be observed in a simple system in case of a periodic forcing\cite{Basak2024partial}. 
In case of a chaotic attractor, the system attains probabilistic properties, so that the corresponding critical transition is better described by a distribution of tipping probabilities\cite{Kaszas2019} rather than by a single deterministic value. A similar type of rate dependence can also be observed in stochastic dynamical systems~\cite{Lohmann2021}.

Note that, in case of a periodic or chaotic attractor, whether the system tips or not can depend on its initial position on the attractor. This dependence on the initial conditions may also lead to a distribution of tipping thresholds and hence can also be thought of as partial tipping\cite{Kaszas2019,Basak2024partial}.
 }

{\color{black} Modern quantitative approaches to reveal the critical rate of R-tipping go beyond the classical bifurcation theory of autonomous systems. Generally, they apply the notion of the (local) pullback attractor and consider bifurcations of such attractors, when the rate of external forcing is being gradually increased. The pullback attractor is an extension of the classical notion of the attractor in autonomous systems, and it is understood as an attracting time-depending set to which the system has evolved at time $t=t_0$, starting from some initial conditions from long past \cite{anagnostopoulou2023nonautonomous}. Bifurcations of pullback attractors, occurring for example via collision with pullback repellors, should provide the condition for R-tipping \cite{hoyer2021rethinking}. Currently, the most elaborated methods deal with the scenario where the time dependence in the parameter (external forcing) is asymptotically constant: it has finite limits when $t\rightarrow \pm \infty$, thus the external forcing remains globally bounded \cite{ashwin2017parameter}. In this case, one can use the method of compactification \cite{wieczorek2023rate}. The initial nonautonomous system is transformed into the so-called augmented system, where time is considered to be an extra dynamical variable. The time is then transformed in a way to make this additional variable bounded (mathematically, this can be done for exponentially asymptotically constant time dependence of the parameter). This gives an autonomous system with bounded solution. Finally, the obtained compactified system can be analyzed using the exiting methods for autonomous systems, for example, singular perturbation methods for fast-slow systems \cite{ashwin2017parameter}. Other methods to predict the critical rate in R-tipping include:  Melnikov-like methods (using asymptotic series expansions) \cite{kuehn2022estimating},  finite-time Lyapunov exponents \cite{hoang2016finite}, construction of safe regions around moving stable equilibria, or various geometric singular perturbations methods \cite{mitry2013excitable,vanselow2019very}. Note that the existing methods are mostly focused on the scenario, when the moving states (pullback attractors) are equilibria, whereas analytical methods to estimate the critical rate of P-tipping, involving moving limit cycles or more complicated attractors, has much less developed \cite{ritchie2023rate}. Also, we should say that ecosystem modelling, the scenario, assuming asymptotically constant external forcing, might not be appropriate, as for example, in the case of a gradual raise of the global temperature. Therefore, novel methods, as those considering locally bounded (but not necessarily asymptotically constant) parameter change need to be developed \cite{hoyer2021rethinking}.}

{\color{black}We also mention that, in a nonautonomous system, the two generic tipping scenarios described above, i.e.~tipping when a parameter crosses its critical value and tipping when the rate of change exceeds a critical value, can interact in a highly nontrivial way\cite{OKeeffe2020}, leading to a more complicated structure of the parameter space and opening a possibility of tipping prevention. In particular, if the tipping point overshooting time is short compared to the intrinsic system's timescales, tipping can be avoided through a parameter shift reversal\cite{OKeeffe2020,Ritchie2021overshooting}.}

The classical B-tipping mechanism does not require any stochasticity. However, stochasticity (to which we refer here as noise) is ubiquitous in all natural systems. Once  noise is present, under some weak conditions on the properties of noise , a regime shift in a bistable or multistable system can occur (known as noise-induced tipping or N-tipping) without any parameter change, as noise can occasionally push the system out of its stable basin to the basin of another attractor  (Figure \ref{Figure2}b). Note that this tipping scenario does not require any change in the system's parameters (meaning no change in the environmental conditions), as noise can act only on the state variables of the system. Correspondingly, since the effect of noise does not destroy the original stable steady state, a reverse transition may also be possible. A life-time of the ecosystem in each of the stable steady states (more generally, in each of the stable dynamical regimes) is therefore finite. On a longer timescale, the corresponding ecosystem dynamics becomes a succession of alternating steady states (or regimes): the phenomenon that is sometimes referred to as metastability \cite{margazoglou2021dynamical}. {\color{black}When proposing our classification of tipping scenarios, we should say that a strict  borderline between R-tipping and N-tipping mechanisms is hardly possible, for example, in the case of highly correlated noise, as with the red noise \cite{morr2024detection}. Indeed, such a model would  contain both effects of stochasticity (N-tipping) and will also depend on a variable parameter (R-tipping).}

The notion of quasi-potential is an essential ingredient for performing a quantitative analysis of the metastability of a stochastically forced system. As a paradigmatic example, let us follow the Hasselmann program \cite{Hasselmann1976,LucariniChekroun2023} and consider the dynamics described by the following Ito {\color{black}stochastic differential equation (SDE)\cite{Pavliotis2014}: 
\begin{equation}
\mathrm{d}\mathbf{x}=\mathbf{f}(\mathbf{x},\nu)\mathrm{d}t+\sigma\Sigma(\mathbf{x})\mathrm{d}\mathbf{W},\label{Ito}
\end{equation}
where $\mathbf{x}\in\mathbb{R}^N$,$\mathbf{f}:\mathbb{R}^N\rightarrow \mathbb{R}^N$ is a smooth drift, $\sigma>0$ is the noise intensity, $\nu$, as above, is a parameter,$\mathrm{d}\mathbf{W}$ is a vector whose components are N independent increments of Gaussian processes, and $\Sigma(\mathbf{x}):\mathbb{R}^{N\times N}\rightarrow \mathbb{R}^N$ defines the noise law. We assume that the noise is non-degenerate, which boils down to imposing that $\Sigma(\mathbf{x})\Sigma(\mathbf{x})^T$ is invertible everywhere. 

The evolution of the probability density of the system is described by the following Fokker-Planck equation \cite{Pavliotis2014}:
\begin{equation}
\frac{\partial p(\mathbf{x},t)}{\partial t}=\mathcal{L}_\nu p(\mathbf{x},t) \quad \mathcal{L}_\nu \bullet=  -\partial_i(\mathbf{f}(\mathbf{x},\nu)_i\bullet,) + \frac{\sigma^2}{2} \frac{\partial^2}{\partial_i\partial_j} \left( D_{ij}(\mathbf{x}) \bullet \right)\quad \boldsymbol{D}=\boldsymbol{\Sigma}\boldsymbol{\Sigma}^T\label{FP}
\end{equation}
The invariant measure of the system $\rho(\mathbf{x},\nu)$  describe its steady-state statistics and obeys the following equation $\mathcal{L}_\nu\rho(\mathbf{x},\nu)=0$. Under rather general hypotheses, in the weak noise limit $\sigma\rightarrow 0$ the invariant measure of the system can be approximated as $\rho(\mathbf{x},\nu)\sim Z(\mathbf{x},\nu)\exp(-2\Phi(\mathbf{x},\nu)/\sigma^2)$, where $\Phi(\mathbf{x},\nu)$ is the so-called quasi-potential \cite{Graham1991,zhou2012,LucariniBodai2020}. 

The function $\Phi$ is the generalization of the energy landscape for general nonequilibrium systems.} 
{\color{black}Local minima of $\Phi$ are associated with attractors of the dynamics. If multiple competing states are present, $\Phi$ also provides information on the properties of the boundaries between the competing basins of attraction. Indeed, $\Phi$ is a saddle at the edge states. Note that if we consider smaller values of $\sigma$, the probability density concentrates in the local minima of $\Phi$. Note also that if for a critical value of the parameter $\nu=h$ we have B-tipping, the attractor and the edge state collide and, similarly to the equilibrium case discussed above, the corresponding minimum and saddle of $\Phi$ merge. Note that the accurate treatment of the simultaneous limits $\nu\rightarrow h$ and $\sigma\rightarrow0$ is very nontrivial as B- and N-tipping processes interfere and gain relative prominence depending on such limits are realised. Note also that the presence of noise can strongly interfere with the process of R-tipping by lowering the value of the critical rate of change of the parameter that is conducive to tipping \cite{Slyman2023}.
Metastability can occur on multiple scales and can be organised in a hierarchical manner. This is then reflected in the properties of the quasi-potential $\Phi$, see \cite{margazoglou2021dynamical} for a detailed discussion.}


In the weak-noise limit, transitions between competing metastable states occur predominantly along special trajectories - the instantons - that connect an attractor and a nearby edge. Instantons can be obtained by solving a variational problem associated with the Friedlin-Wentzell action \cite{Freidlin2012}. In the case of nonequilibrium systems, {\color{black}spontaneous relaxation trajectories (i.e heteroclinic orbits connecting an  edge and an attractor) and instantonic trajectories  are geometrically different } \cite{Graham1991,LucariniBodai2020}. Note that if we consider a weak but not infinitesimal noise, the most likely as well as the mean escape paths are better captured by solving a modified variational problem based on the Onsager-Machlup functional \cite{Fleurantin2023,Borner2024}.
Finally, the average permanence time of the system in the vicinity of a reference state follows the scaling $\log(\tau)\sim(2\Delta \Phi/\sigma^2)$, where $\Delta \Phi$ is the lowest quasi-potential barrier confining the reference state \cite{Galfi2021}. 

A dynamical system can also be pushed out of the basin of its stable equilibrium by a single short-term perturbation - a `shock' - provided its magnitude is sufficiently large\cite{halekotte2020}. This mechanism is known as shock-tipping or S-tipping. Probably the most remarkable example of this tipping scenario occurred in the Earth deep history when Earth collided with a large bolide about 65 Ma ago; that singular event led to a dramatic global climate change, in particular resulting in a mass extinction\cite{alvarez1980}. In the present-day Earth, S-tipping becomes increasingly relevant due to the increased frequency and severity of extreme climate events\cite{romanou2024}. 
{\color{black}
In a more general case, there may be more than one shock. In that case, if the interval between the shocks is sufficiently small, e.g.~smaller than a characteristic system's relaxation time, the effect of shocks can be mutually amplifying: while each individual one might be under-critical (not leading to tipping), a succession of them can make the system to tip~\cite{Basak2024}.}

\begin{table}[!b]
 \small 
\centering
\caption{Different types of tipping points and tipping mechanisms.}
\begin{tabular}{|p{2.8cm}|p{2.3cm}|p{7.0cm}|p{3.5cm}|}

\hline 
{\bf Preconditions} & {\bf Tipping type} & {\bf Brief description} & {\bf Examples}\\
\hline 
Change in parameter values (e.g.~a persistent environmental change or trend) & B-tipping (bifurcation tipping) & Transition of a dynamical system to an alternative regime (e.g., a different steady state or oscillations with different frequency or magnitude) 
 & Amazon forest\cite{scheffer2001}, savanna vegetation\cite{rietkerk2004} \\
\hline 
Sufficiently fast (overcritical) rate of parameter change & R-tipping (rate induced tipping) & Tipping occurs if the rate of change in a certain parameter(s) exceeds a critical value & ``Compost bomb''\cite{Wieczorek2011compost,Clarke2021compost}, r-tipping in bacterial communities, plankton blooms, complex mutualistic networks\cite{ashwin2012tipping,karita2022,osullivan2023,ritchie2023rate,panahi2023,vanselow2024}\\
\hline 
Presence of noise acting on state variables & N-tipping (noise induced tipping) & Transition of a system to an alternative regime due to the impact of noise. Increase in the noise variance {\color{black}(noise strength)} may increase the frequency of tipping & Common in all natural systems with alternative steady states\cite{scheffer2012}\\
\hline 
Singular large perturbation of a state variable & S-tipping (shock tipping) & Transition to an alternative steady state or regime following a single large perturbation of state variables & Extreme/rare weather events\cite{romanou2024}, bolide-Earth  collision\cite{alvarez1980}\\ 
\hline 
Non-Gaussian, `anomalous' noise (e.g.~Levy-type) & {\color{black}AN}-tipping  & Transition to an alternative steady state or regime, but with a distribution of exit times qualitatively different from N-tipping & Amazon forest\cite{serdukova2017}, vegetation systems more generally\cite{zhang2023}, weather\cite{yang2022} \\ 
\hline 
Periodic or quasi-periodic dynamics & P-tipping (phase tipping) & Similar to N and S tipping but showing a much higher probability of tipping at certain specific times (or ``phases'') due to the effect of noise. Increase in the noise variance may increase the likelihood of tipping & Cyclic ecosystems\cite{alkhayuon2021}\\
\hline 
Ghost attractors; saddle points; {\color{black}slow manifolds in systems with} multiple timescales & LT-tipping (tipping at the end of long transients) & For systems exhibiting a “long transient” (LT): transient dynamics that mimics a self-sustained (asymptotic) regime but only lasts for a finite time & Inherent in many ecological and other environmental  systems\cite{hastings2018transient,morozov2020,hastings2021,morozov2024long,Mehling2024,borner2025}\\
\hline 
\end{tabular}
\label{table1}
\end{table}	

Note that the timing and  magnitude of the shock(s) can usually be considered as random variables. 
One can therefore regard the S-tipping as a particular case of N-tipping where the noise has special special properties. 
It a more general case, many small frequent random perturbations are combined with rare large ones. Remarkably, this is exactly the case of `fat-tailed noise' described by a power law\cite{mandelbrot1982,shlesinger1993}. An important example of such heavy-tailed, impulsive processes is given by Levy noise, which can be understood in a coarse-grained sense as compound Poisson process, whereby singular perturbations arrive at random times. Tipping due to the effect of such `anomalous noise' 
{\color{black}
was shown to exhibit properties considerably different from standard N-tipping in the case of a Gaussian noise\cite{serdukova2017,lucarini2022}. Indeed, in case of a fat-tailed noise the whole notion of `small noise', which is central is many studies, becomes pointless as the probability of a large kick over any finite time interval is not small. 
In the case of Levy processes the whole notion of quasi-potential becomes immaterial, whilst transitions occur when the displacement due to the occurrence of a singular perturbation kicks the system into the basin of attraction of the competing state, as defined in the absence of noise. 
In order to emphasize these qualitative differences, here we refer to tipping due to the effect of a fat-tailed noise as AN-tipping. 
A more general} theoretical framework for tipping due to the effect of stochastic factors should consider a combination of Levy noise and Gaussian noise, see \cite{Chao2019,Dai2022,Chekroun2025}.

Many ecosystems and ecological communities exhibit cyclic dynamics where the cycles can be either self-organized (cf. prey-predator cycles) or externally forced (e.g. by seasonal changes). In such a case, the  sensitivity of the system to perturbations can depend on its position (“phase”) in the cycle, and thus on the timing of the perturbation. For instance, a noise of the same intensity may lead to tipping when applied at one time during the cycle (e.g. if/when the system trajectory is close to the boundary of the attraction basin), but will not lead to tipping at another time. A similar statement is also true  for a single perturbation of a given magnitude. This type of tipping due to a system's time-varying (phase dependent) sensitivity is called P-tipping\cite{alkhayuon2021}. 

Finally, there is one more tipping mechanism that acts somehow differently from those described above. A dynamical system can be in an apparently self-sustained stable regime mimicking an asymptotic regime over a long time before ‘suddenly’ exhibiting a regime shift. The essential difference from other mechanisms is that such regime shift or tipping is not preceded by any change in parameter values. Else, it can follow a parameter change that happened a very long time before the tipping. We refer to such a regime shift that occurs at the end of long transient dynamics\cite{hastings2018transient} as {\it long-transient} tipping or LT-tipping. Note that, as the existence of long transients is linked to some common dynamical system structures (ghost attractors, unstable manifolds, multiple timescales, etc.)\cite{morozov2020}, neither the presence of any noise nor a single large perturbation are  required for LT-tipping to happen. Long transients and the corresponding LT-tipping have been observed in many real-world ecological and environmental systems\cite{morozov2024long}.

\section*{Cascading tipping} 

\begin{figure}[!b]
\centering
\subfigure[]{\includegraphics[scale=0.11,angle=0]{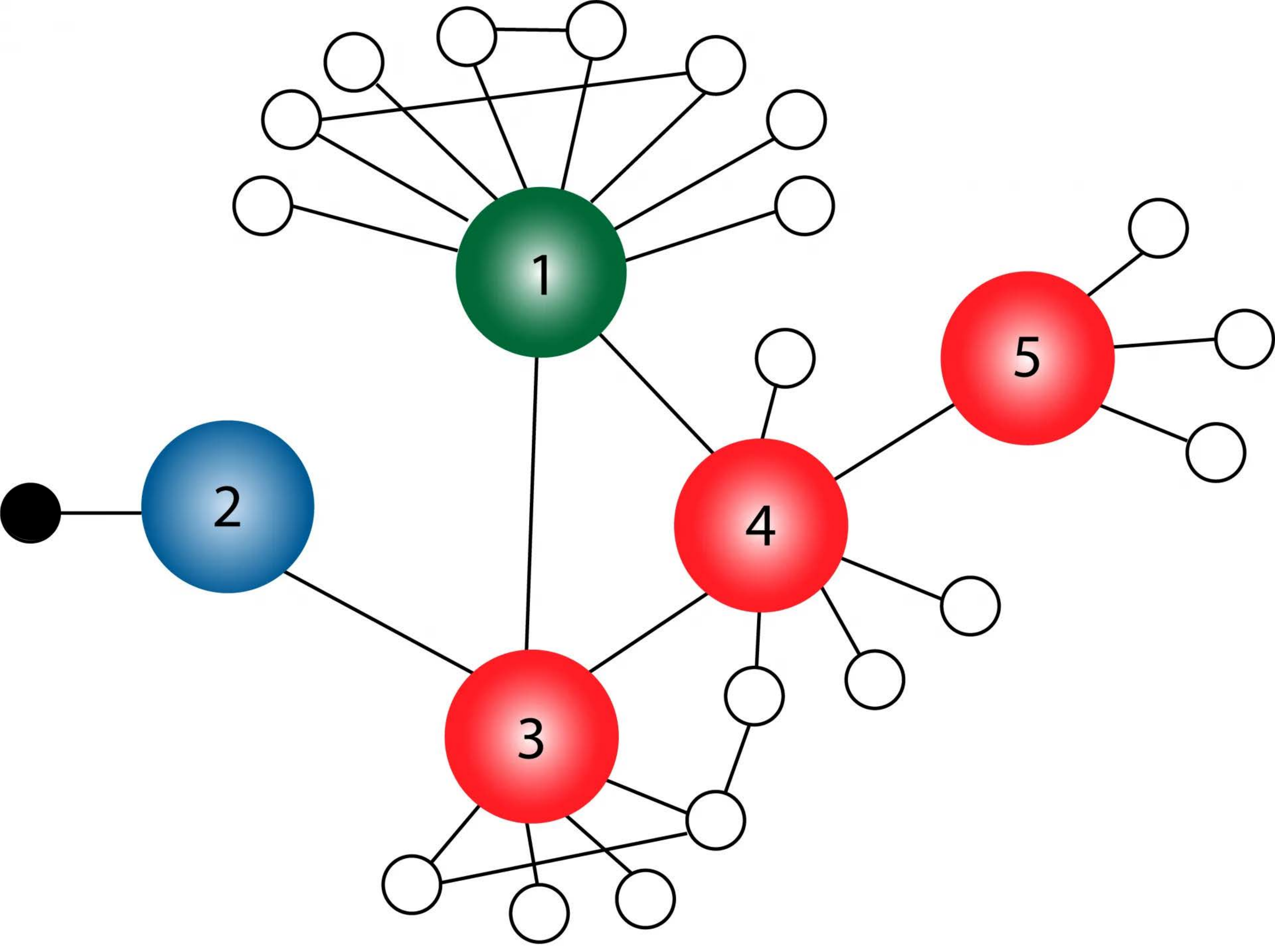}}\hspace{8mm}
\subfigure[]{\includegraphics[scale=0.46,angle=0]{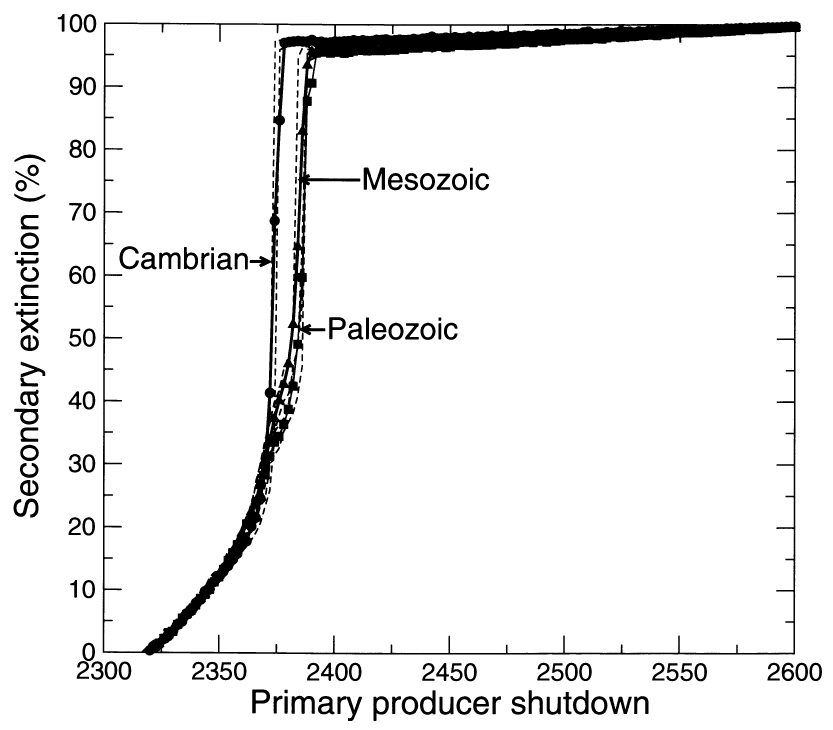}}
\caption{\small Cascading tipping on trophic networks. (a) An example of heterogeneous foodweb where tipping (species extinctions) at nodes 1-5 leads to tipping in other parts of the web. Figure adapted from\cite{palmer2023}. (b) Effect of tipping cascade in an ancient food web\cite{Roopnarine2006} resulting in a highly nonlinear, threshold-type response of the ecosystem to the initial tipping in parts of the web. Figure adapted from\cite{Roopnarine2006}.}
 \label{Figure4netwcasc}
\end{figure}

Due to its inherent structural and spatial complexity, an ecosystem often can be considered as a number of subsystems coupled among themselves. This leads to a possibility of cascading tipping (or a tipping cascade) when tipping in one subsystem, e.g.~species extinctions, triggers tipping in other subsystems through a sort of chain reaction\cite{basak2025cascade,wunderling2020interacting,klose2021we}. This can happen in  several different contexts. One of them is the heterogeneity of food webs and the existence of bottlenecks and key species; a schematic example is shown in Fig.~\ref{Figure4netwcasc}a. Tipping in any of the bottleneck subsystems (e.g.~nodes 1-5 in Fig.~\ref{Figure4netwcasc}a) would likely lead to tipping in other connected subsystems\cite{pace1999,dunne2009}; this is called secondary extinction (see Fig.~\ref{Figure4netwcasc}b). The extent of such a `domino effect' and/or the magnitude of the resulting impact on the ecosystem as a whole depends on the position of the bottleneck in the food web\cite{dunne2002}. Extinction of primary producers (cf.~subsystem 2 in Fig.~\ref{Figure4netwcasc}a) known as `bottom-up' cascade would intuitively lead to tipping/extinction in many other subsystems at higher trophic levels. Similar cascading effects have also been reported in physical and engineered systems, including power grid failures where collapsing of the first element can result in subsequent numerous failures of other connected nodes \cite{gao2012networks}. Much less intuitively, due to highly nonlinear interactions between  species in a foodweb, a `top-down' cascade\cite{HSS1960} can also occur\cite{Heath2014,Roopnarine2006}. Tipping at a high trophic level (e.g.~extinction of a top predator) may lead to tipping at low trophic levels of the foodweb by releasing pressure on intermediate consumers and hence causing an over-exploitation of a food resource\cite{Heath2014,Roopnarine2006}. 

Note that cascading tipping in ecosystems may occur on networks of a different origin. Due to its spatial structural complexity, an ecosystem often consists of a fragmented habitat where the same species exists at different separated spatial locations coupled by dispersal. Species extinction in one location or a collapse of one local community may then lead to a collapse of the entire dispersal network and thus the whole ecosystem\cite{Staddon2010,Reji2025}. 

In Earth systems more generally, where ecosystems are coupled to climate\cite{lenton2008tipping,Boers2017,wunderling2024climate} as well to social and economic systems\cite{Lenton2013,watts2002simple}, tipping cascades may follow a variety of even more complicated scenarios.
We firstly consider cascades in fully deterministic systems. One tipping scenario occurs in master-slave {\color{black}(known also as driver-driven)} systems where the leader subsystem ( systems`master') exerts control over the follower subsystem (`slave'), which in turn can regulate another slave system \cite{brummitt2015coupled, basak2025cascade}. We assume that each isolated subsystem (a tipping element) within this unidirectional chain has at least two qualitatively different coexisting states. In case each subsystem in the chain is close to its tipping threshold, tipping in the master system would propagate down the chain resulting in the domino effect \cite{wunderling2020interacting,klose2021we}.  An example of such cascade tipping {\color{black}via the master-slave mechanism is the observed regime shift} between eutrophic and oligotrophic states in a connected cascade of freshwater lakes \cite{hilt2011abrupt}. Interestingly, even in a simple unidirectional chain, the effect of cascading hopping can occur where propagation of tipping across the system may hop over some subsystems if they are characterized by higher tipping thresholds. Such subsystems do not tip themselves yet trigger tipping in their neighbor systems down the chain\cite{brummitt2015coupled,basak2025cascade}. {\color{black}An example of cascade hopping in a complex socio-economic-environmental system that included 12 countries is the propagation of protests during the Arab spring (2010-2011). In this system, sudden massive street protests (understood as tipping) started locally but then propagated across a large geographical region by hoping over some countries \cite{brummitt2015coupled}.}  

{\color{black}In a master-slave chain, also an \textit{indirect tipping} can happen\cite{basak2025cascade} where tipping does not occur in the master system which is subject to the parameter drift but instead takes place in the slave subsystem. In a master-slave chain with three subsystems, more complex combinations of cascading tipping and indirect tipping can occur, as the parametric drift occurring in the master subsystem can cause tipping in either of the two slave systems down the chain\cite{basak2025cascade}.}
A more complicated scenario involves bidirectional connections, where subsystems affect each other via positive or negative coupling \cite{klose2021we}. 



Interacting subsystems are often connected via a complex network and the nature of this network plays a crucial role in the occurrence and propagation of tipping cascades \cite{kronke2020dynamics}. 
{\color{black}The domino effect on a complex network has been reported in the study on the Amazon rain forest degradation} where coupling between local patches is due to moisture-recycling mechanisms \cite{wunderling2022recurrent}.  {\color{black} Another empirical study shows a domino effect in a large spatial aquatic system in the western Baltic Sea. Propagation of a regime shift through space was observed from the open sea towards mainland, resulting in the change of the dominance from large predatory fish to smaller prey fish \cite{eklof2020spatial}}. Importantly, the realization of tipping cascade on a network can depend on which subsystem has initially tipped \cite{wunderling2020motifs,wunderling2021modelling}.  

\begin{figure}[!b]
\centering
\includegraphics[scale=0.45]{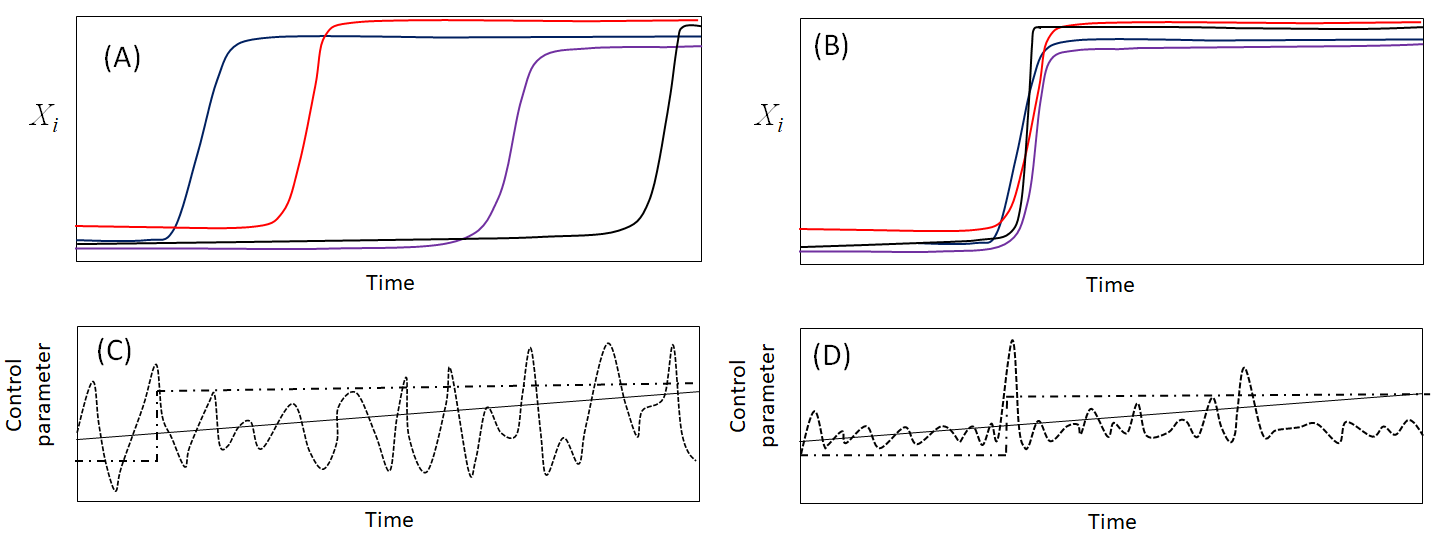}
\caption{Different types of cascading tipping. (A) Dynamical pattern of cascading tipping in a system of coupled bi-stable subsystems $X_i(t)$ on a network. Initially, all subsystems stay in the lower (non-tipped) state; then further tipping of each subsystem occurs as a separate event. For the same observed pattern shown in (A), the mechanism of cascading can be different (quasi-cascading; multi-phase cascading; slow domino effect, noise-induced cascading without domino effect, etc), and this depends on the nature of the variation of the control parameter. (B) Joint cascading tipping (joint tipping/fast domino effect): regime shifts of $X_i$ occur simultaneously. (C)-(D) Scenarios of variation of the control parameter, resulting in cascading in shown (A) and (C), respectively: a gradual increase (solid line); stochastic variation (dashed line), {\color{black} and step-wise abrupt switch (dashed-and-dotted line).} }
\label{figure_cascades}
\end{figure}

{\color{black} 
We emphasize that tipping in one subsystem (e.g.~as a result of an increase in external forcing) is not by itself sufficient to initiate a tipping cascade, in particular if the coupling between the subsystems is weak. However, in case an increase in the forcing continues, it may cause other  subsystems tip {\it before} their own tipping points are crossed, i.e.~earlier than it would happen for each of the subsystems in isolation. In such case, the change of the internal state in the subsystem that tips first effectively works as an addition to the external forcing. Note that, depending on the rate of change in the external forcing, the interval between the tipping of the first subsystem and tipping of other subsystems can be long. This is called a two-phase tipping  \cite{klose2021we} or a multi-phase tipping in case of multiple tipping elements.
A well known example of two-phase cascading in a climate system is given by the interaction between the AMOC and the El Ni\~{n}o Southern Oscillation (ENSO) \cite{dekker2018cascading,klose2021we}. 
An example of multi-phase cascading is given by the Arab spring where tipping occurred in some semi-isolated countries (that were only weakly coupled to the rest of the system) because arguably the tipping across the region was triggered by a change in some global factors (e.g.~rising food prices across the entire geographic region\cite{brummitt2015coupled}).}  Finally, we mention that tipping cascade can also be triggered if the parameter change is fast enough; this is known as a rate-induced tipping cascade \cite{klose2024rate}.




A key aspect of multi-phase cascades is the characteristic timescales of tipping elements\cite{Ritchie2025arxiv}. 
As a prototypical example, consider a unidirectional master-slave chain containing interconnected subsystems $A, B, C$, etc. Let the state before tipping be $(A,B,C,\ldots)$. An initial collapse of $A$ can trigger a domino effect causing the subsequent collapse of $B$, which in turn will result in the collapse of $C$, eventually leading to the final `tipped' state $(A^*,B^*,C^*,\ldots)$. In case the internal characteristic timescale of subsystem $B$ is long, after subsystem $A$ tips, subsystem $B$ will still remain in its original state for some (possibly very long) time, so that a transient `metastable' state $(A^*,B,C,\ldots)$ emerges \cite{ashwin2025early}. 
The lifespan of the metastable state(s) can be made longer by the presence of long transients caused by ghost attractors and crawl-bys (saddles) \cite{hastings2018transient,morozov2024long}. However, in case the internal timescales of subsystems $B$ and $C$ are short, all three subsystems tip at approximately the same time, so that a fast transition to the final state $(A^*,B^*,C^*,\ldots)$ occurs. This scenario is called a joint cascading \cite{klose2021we} (also known as a fast or synchronized domino effect\cite{ashwin2017fast}). 
The difference between the domino cascading and the joint cascading is schematically shown in Figs.~\ref{figure_cascades}A and \ref{figure_cascades}B {\color{black} (where the corresponding scenarios of change of the control parameter is shown in the bottom panels). 
An empirical example of joint cascading can be the reported degradation of the Caribbean coral reefs with a quick loss of about 50\% of the coral cover \cite{cooper2020regime}}. 

{\color{black} 
Another scenario of a tipping cascade may occur in a chain or network of subsystems in case they are situated along an environmental gradient, so that their tipping thresholds are arranged in a hierarchical order accordingly. In this case, an increase in the global external forcing (e.g.~temperature increase) would cause a sequence of tipping events along the chain in a manner apparently similar to the domino effect. An important difference however is that, in this scenario (to which we refer as quasi-cascading), tipping in neighboring subsystems is not necessarily caused by direct influence of tipping in the adjacent subsystems: in fact, it can occur in a system of uncoupled elements or when coupling is weak \cite{dekker2018cascading}.}

Including environmental and demographic stochasticity can modify the above scenarios and the classification of tipping cascades. Noise can trigger tipping cascades that would not be possible in the corresponding deterministic system. The concepts of multi-phase cascading and domino effect may become hard to distinguish in the presence of noise. 
For a slow domino effect, noise-induced tipping results in a regime shift of one subsystem, while the other subsystems can remain 
at their current state for a long transient time \cite{ashwin2017fast,creaser2018sequential}. 
{\color{black} However, noise can reduce the transient time of the domino effect, as compared to the corresponding deterministic system} \cite{ashwin2017fast}. 
In particular, a spatially correlated noise can synchronize the dynamics of separate tipping elements hence turning the domino effect to joint cascading.


The existing mathematical theory of tipping cascades is mostly limited to networks of bistable elements, in many cases containing only 2-3 subsystems with unidirectional coupling \cite{ashwin2017fast,klose2021we}. This is because of the complexity of the system, where adding just one extra tipping element to the network or {\color{black}considering more complicated coupling (e.g.~a bump-shaped coupling or nonlocal/nonlinear coupling with its strength dependent on the state of the adjacent subsystems)} may bring a new type of tipping \cite{basak2025cascade,ashwin2025early}. 
Modeling cascading effects in real-world Earth systems that include multiple interacting biological and non-biological subsystems is generally extremely challenging \cite{wunderling2024climate}. A typical current approach consists of constructing a `master model' that includes a large number of subsystems. The model is then studied via extensive simulations trying to connect potential tipping events in individual subsystems. However, such a `brute force' approach is often lacking in understanding of the generic mechanisms behind the tipping\cite{d2020linking,wunderling2024climate}. For example, it may  be difficult to conclude which particular type of cascading scenario is realized \cite{wunderling2020interacting}. {\color{black}This is schematically demonstrated  in Fig. \ref{figure_cascades}, where the same cascading patterns (shown in Figs.~\ref{figure_cascades}A and \ref{figure_cascades}B) can be potentially caused by different types of variation of the parameter (shown in Figs.~\ref{figure_cascades}C and \ref{figure_cascades}D respectively), such as a gradual increase; stochastic oscillation, and a step-wise abrupt change.} 


\section*{Early warning signals - progress, challenges and open questions}


Anticipating tipping points due to various mechanisms is of key importance in Earth sciences, and the literature on early warning signals (EWS) of critical transitions is abundant. In particular, it was reported that many different systems show similar behaviour when they approach a critical transition: their variance and autocorrelation increase \cite{scheffer2009early}. Most  existing methods of EWS are summarised in a recent review by \cite{dakos2024tipping}. The classical methods of EWS are based on the paradigmatic concept of critical slowing down (CSD). This concept states that near a bifurcation point, the recovery rate of the system progressively deteriorates; therefore, to detect possible tipping point, one needs to follow the key indicators, such as the variance, temporal/spatial correlation, and recovery rate. However, classical CSD-based indicators of regime shift often fail to provide reliable forecasts. For example, they are unable to reveal the regime shift even while an abrupt transition  occurs in the system \cite{burthe2016early,dakos2024tipping}. Another possible reason for the failure of the CSD-based methods is the effect of noise.
{\color{black}Indeed, in the case where the confining quasi-potential (cf.~Fig.~\ref{Figure1}) is shallow, 
the destabilising effect of stochasticity may be not compensated by the system attraction to the stable steady state,
so that even the slightest amount of noise (or of extra active forcing) would cause the system to tip\cite{Budd2026criticalnoise}.} 

{\color{black}It is important to note that whilst CSD is conceptually linked to bifurcation-induced transitions, in actual systems its common EWS can exhibit  signals that mimic true CSD without being caused by an approaching tipping point \cite{Boers2025}. As an example, both variance and autocorrelation may increase in purely linear systems simply due to non-stationary correlated noise \cite{Boers2021}.}

{\color{black}Applying the formalism of the Koopman/Kolmogorov operator \cite{Budisic2012} or, equivalently, of the Perron/Frobenius operator \cite{B00}, it is possible to greatly generalise  \cite{ChekrounJSPI,GutierrezLucarini2022} the classical CSD arguments for systems whose dynamics is described by Eq. \ref{Ito}. The two approaches are equivalent as the first one focuses on the evolution of observables, and the second one on the evolution of probability measures. We recall here some key results, whilst a more detailed analysis can be found in \cite{ChekrounJSPI,LucariniChekroun2023,Chekroun2025}. Under rather general conditions, the operator $\mathcal{L}_\nu$ introduced in Eq. \ref{Ito} admits eigenvalue/eigenvectors pairs $(\lambda_j,\rho_j)$ such that $\mathcal{L}_\nu\rho_j=\lambda_j\rho_j$. The $\lambda_j$'s have nonpositive real part, where $\lambda_0=0$ is the eigenvalue corresponding to the invariant measure, as discussed above. The eigenvalues $\lambda_j$'s $ \in \mathbb{C}$ can be hierarchically ordered as $\lambda_0 = 0 > \Re(\lambda_1) \geq \Re(\lambda_2) \geq \dots $ and  represent intrinsic relaxation rates of the ensembles shaping the statistical dynamical properties of the system associated with the Perron-Frobenius modes $\rho_j$'s. 

Equivalently, one can instead study the eigenpairs $(\lambda_j,\phi_j)$ such that $\mathcal{L}^T_\nu\phi_j=\lambda_j\phi_j$, where $\mathcal{L}^T_\nu$ is the adjoint of the operator $\mathcal{L}_\nu$ introduced in Eq. \ref{Ito}. In this case, the $\phi_j$'s are the Koopman/Kolmogorov modes. } 

{\color{black}The discrete, isolated, eigenvalues $\{ \lambda_j \}_{j=1}^N$, with $N$ possibly infinite, are sometimes called (stochastic) Ruelle-Pollicott resonances \cite{ChekrounJSPI}. The so-called continuous spectrum describes the very fast decaying modes. See Fig. \ref{Figure4} for a schematic depiction of the spectral properties of $\mathcal{L}_\nu$.}

{\color{black}As shown in \cite{LucariniChekroun2023}, it is possible to approximate the lagged autocorrelation of a general observable $\Psi$ as: 
\begin{align}
\label{eq: spectral decomposition corr and Green's functions}
    C_{\Psi,\Psi}(t) &= \sum_{j=1}^N \alpha_j(\Psi)e^{\lambda_j t},
\end{align}
where the formula for the coefficients $\alpha_j(\Psi)$ can be found in \cite{LucariniChekroun2023}. The essential result is that no matter the choice of the observable, the correlation is written in terms of dictionaries of given exponential functions \cite{GutierrezLucarini2022,LucariniChekroun2023}. 
{\color{black}
\begin{figure}[!b]
\centering
{\includegraphics[scale=0.5]{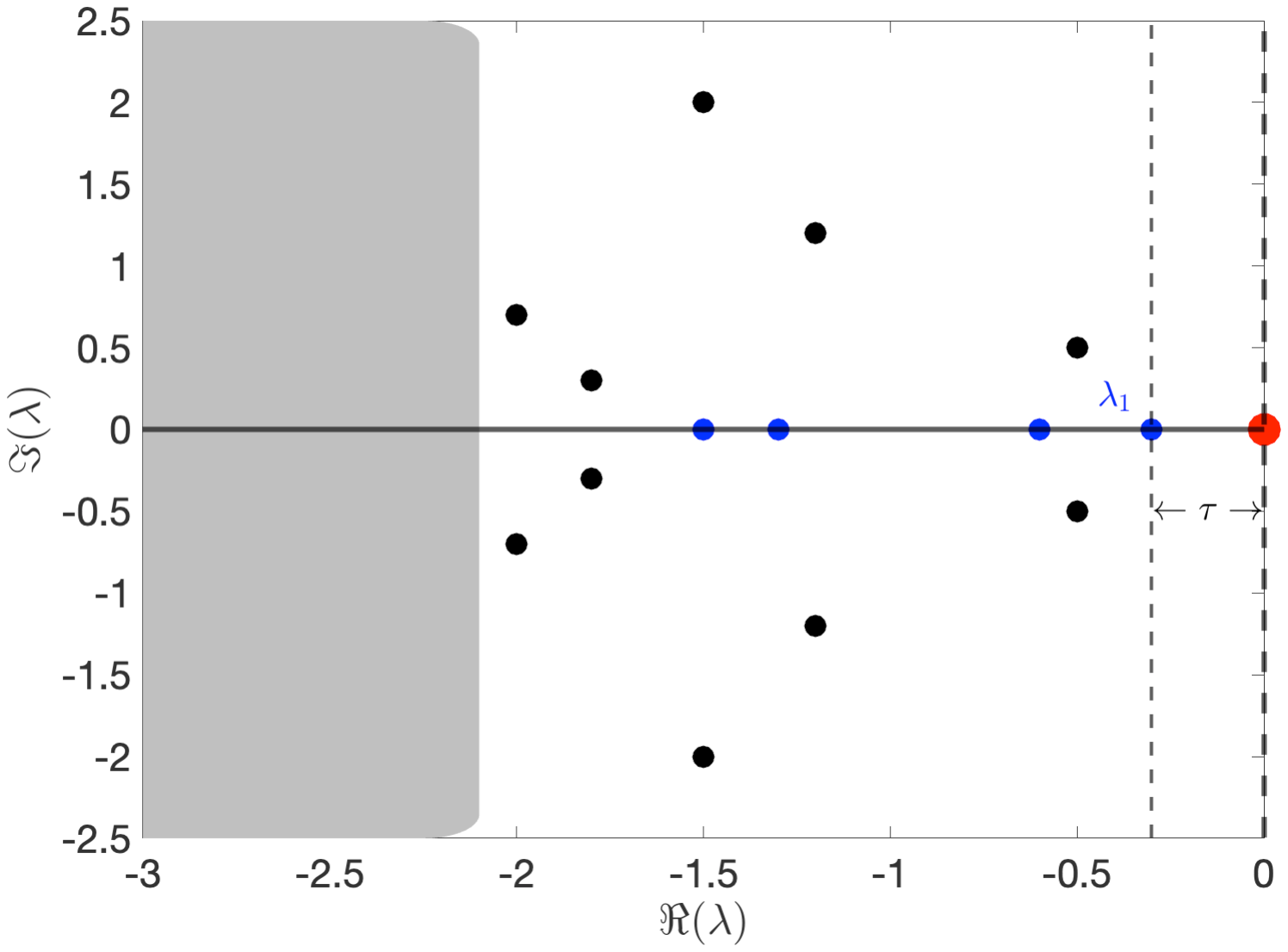}}
\caption{\small Schematics of the eigenvalues of the generator of the Kolmogorov operator with indication of the spectral gap $\tau$. Isolated spectrum: the blue dots indicate the real eigenvalues, whilst the black dots indicate the complex conjugate pairs. The continuous spectrum is indicated in grey. Tipping behaviour occurs when, as $\nu\rightarrow h$, the spectral gap shrinks to zero. See text for details.}
 \label{Figure4}
\end{figure}
}

Tipping behaviour occurring as $\nu\rightarrow h$ is associated with the real part of the subdominant Ruelle-Pollicott eigenvalue  approaching zero (or, in other terms, the spectral gap $\tau=|\Re(\lambda_1)|$ shrinking to zero), so that the corresponding characteristic time scale of decay diverges, see Fig. \ref{Figure4}. If $\tau$ shrinks for a pair of eigenvalues, we will have at the transition a critical oscillation, which is reminiscent of a Hopf bifurcation but occurring in potentially very high dimension \cite{ChekrounJSPI,borner2025}. 
The mode $\rho_1$ associated with the subdominant $\lambda_1$ (or with the pairs of critical complex conjugate $\lambda$'s) define the so-called critical mode of variability, which can be identified with the so-called degenerate fingerprinting \cite{Held2004}. Such a mode defines the precursor of the critical transition. The lagged correlation $C_{\Psi,\Psi}(t)$ can act as an EWS for the tipping behaviour only if the chosen observable has a strong projection on the critical mode. This becomes especially clear if one consider the expression for the integrated correlation time for 
\begin{align}
   \int_0^\infty \mathrm{d}\tau \tau C_{\Psi,\Phi}(\tau) &= \sum_{j=1}^N \frac{\alpha_j(\Psi,\Psi)}{\lambda_j}\label{integrated}
\end{align}
Hence, if one consider systems that are even moderately dimensional, a well-informed choice of the observable is critical for defining an efficient EWS.} Observables that clearly distinguish between the attractor and  the nearby edge state tend to have a very strong projection on such a critical mode, thus providing a targeted early-warning indicator \cite{Lohmann2025}. 

The great advantage of using the Kolmogorov operator formalism is the possibility of taking advantage of it for performing purely data-driven analysis of the system of interest, especially in view of the efficient and powerful (variants of the) extended Dynamical Mode Decomposition \cite{Williams2015a,Colbrook2024}{\color{black}, which can be used to find with great accuracy and efficiency the eigenpairs $(\phi_j,\lambda_j)$.} Although the standard methodology suffers from the curse of dimensionality, its kernelized version \cite{Williams2015b} is adaptive and seems well suited for studying the spectral properties of high-dimensional datasets. The first attempts in this direction are proving extremely encouraging \cite{LUCARINI2026117540} and definitely deserve further attention.

{\color{black}
The Koopman framework is mirrored by algorithms based on deep learning methods, which offer a complementary approach to traditional CSD indicators. Deep learning has proven capable of handling systems with multiple interacting time scales without requiring explicit prior knowledge of the underlying equations. Moreover, these methods can identify the specific dynamical modes or spatial patterns that are most responsible for the impending tipping behavior \cite{Bury2021,Zhuge2025}. This ability to perform both detection and attribution makes deep learning particularly attractive. Furthermore, deep learning methods are showing promising results for anticipating R-tipping behavior \cite{Huang2024}. This suggests that deep learning may help bridge the gap between classical CSD theory, which often assumes slow or quasi-static forcing, and real-world scenarios where forcing rates can be rapid and time-varying.}

Note that cascading tipping poses additional challenges as the autocorrelation of two time series and their interaction need to be analyzed simultaneously. The study \cite{podobnik2008detrended} suggested an altered form of detrended fluctuation analysis (DFA) to assess the cross-correlation between two non-stationary time series and called this method detrended cross-correlation analysis (DCCA). In the computation of the fluctuation function, they used cross-covariance instead of auto-covariance and fitted it to a power law. {\color{black}EWS in the context of indirect tipping poses additional challenges because studying the correlation properties of the master system will bring no signature of the impeding tipping in the slave system\cite{basak2025cascade}.}

{\color{black} The functional analytical framework discussed above is extremely useful for clarifying EWS in the case of cascading tipping and indirect tipping. Assume the following master-slave form for Eq. \ref{Ito}:
\begin{align}
\mathrm{d}\mathbf{x}_1&=\mathbf{f}_1(\mathbf{x}_1,\nu)\mathrm{d}t+\sigma_1\Sigma_1(\mathbf{x}_1)\mathrm{d}\mathbf{W}_1\nonumber\\
\mathrm{d}\mathbf{x}_2&=\mathbf{f}_2(\mathbf{x}_1,\mathbf{x}_2)\mathrm{d}t+\sigma_2\Sigma_2(\mathbf{x}_2)\mathrm{d}\mathbf{W}_2
,\label{Itobis}
\end{align}
where $\mathbf{x}_1\in\mathbb{R}^{M}$, $\mathbf{f}_1:\mathbb{R}^{M+1}\rightarrow\in\mathbb{R}^{M}$, $\mathbf{x}_2\in\mathbb{R}^{N-M}$, $\mathbf{f}_2:\mathbb{R}^{N-M}\rightarrow\mathbb{R}^{N-M}$,  $\sigma_1,\sigma_2>0$ modulate the noise intensity,  $\mathrm{d}\mathbf{W}_1$ ($\mathrm{d}\mathbf{W}_2$) is a vector whose components are M (N-M) independent increments of Gaussian processes, and $\Sigma_{1}(\mathbf{x}_1)\Sigma_{1}(\mathbf{x}_1)^T$ and $\Sigma_{2}(\mathbf{x}_2)\Sigma_{2}(\mathbf{x}_2)^T$ are nonnegative definite matrices. Finally $\nu$ is the parameter of the master system we alter to explore critical behaviour. In this case,  the eigenvalue set expands beyond the master’s spectrum due to the slave’s internal dynamics. The eigenfunctions depend in general on both $\mathbf{x}_1$ and $\mathbf{x}_2$, so no obvious decoupling is realised.

In the case of cascading tipping, one considers the situation where as we near the critical parameter $\nu\rightarrow h$ the master system nears a tipping transition. In this case, it will be possible to find a suitable observable $\Psi=\Psi(\mathbf{x}_1)$ (or, more generically, $\Psi=\Psi(\mathbf{x}_1,\mathbf{x}_2)$) whose lagged correlation can act as an efficient EWS. In the case of indirect tipping, criticality is associated with the dynamics of the slave system. As a result, the projection of critical mode associated with $\nu\rightarrow h$ on an observable depending on $\mathbf{x}_1$ only will be vanishing. Hence, in practical terms, no EWS can be achieved by looking at the master system only. In many applications, this implies the need for a network-aware and/or system-wide monitoring.}

As can be seen, most of the early warning signal approaches outlined here are essentially local in that they apply in the neighborhood of the tipping point.  An alternate approach \cite{xu2023,Xu2021,Xu_etal_2025} based on landscape-flux theory \cite{xu2014potential} breaks the dynamics of a system into a potential and a flux, essentially complementing the idea of the pseudo-potential. Although at this time the approach essentially relies on an analysis of model equations, conceptually the idea could be used to analyze time series to predict tipping points.

In conclusion to this section, we mention that the existing approaches to early warning signals almost exclusively focus on B-tipping mechanism in the presence of small Gaussian noise. 
Correspondingly, the literature on EWS is dominated by discussion of the CSD as the main indicator of approaching tipping point. However, the CSD is not universally applicable\cite{Ditlevsen2010} and there are a number of reasons for that, one of them being that CSD is designed for B-tipping but not necessarily for other tipping mechanisms\cite{lenton2012,Boettiger2013,dakos2015}. 
Extending the theory of EWS onto other types of tipping - in particular, for AN-tipping, P-tipping and LT-tipping (see Table \ref{table1}) - is perhaps the biggest current challenge for anticipating critical transitions.

\section*{Paradigmatic case studies}

The advanced theory of tipping points is a useful approach for understanding a variety of ecological systems \cite{scheffer2001}and we illustrate this by focusing on three separate cases: kelp forest ecosystems \cite{Steneck_etal_2002,Carnell_Keough_2020,Norderhaug_etal_2017, Christie_etal_2024,Ling_etal_2015,Ling_Keane_2024}, coral-algal grazer systems \cite{Holbrook_etal_2016,Mumby_etal_2007,Schmitt_etal_2022,Xu_etal_2025,Sura_etal_2025} and savanna-forest systems \cite{staver2011,Hirota_etal_2011,vanNes_etal_2014}.  In all of these examples, the local in space dynamics consists of transitions among states dominated by different species, with dispersal among different locations also potentially playing a role.  Thus, both the non-spatial versions of tipping, and the spatial versions, show up in understanding these systems on a global scale. 

{\color{black}
We begin with noting that any sufficiently large environmental change inevitably leads to ecosystems tipping in a cascade-like fashion which can be thought of as a master-slave chain. One of the best known cases addresses 
the importance of ocean currents for the marine systems, e.g.~as related to the Atlantic meridional overturning circulation (AMOC). 
The forthcoming AMOC collapse\cite{Weijer2019amoc} is expected to strongly affect marine ecosystems not only locally due to the change in water temperature and the disruption of nutrient supply\cite{Boot2025amoc}, but also globally, through the strong AMOC feedback on global climate\cite{Nian2026collapseAMOC}. 
Understanding of the potential for the AMOC tipping (collapse), 
in particular by focusing on warning signals for the tipping 
is very important\cite{Ditlevsen_etal_2023}, e.g.~based on an analysis extending ideas \cite{Boettiger_Hastings_2012,Boettiger2013}. 
Studies on the observed AMOC weakening (and hence its anticipated tipping) also emphasized the importance of statistical approaches\cite{Ditlevsen_etal_2023}.
  }

Kelp forests, at least for centuries driven by human activities and changing climatic conditions, have undergone transitions between desirable states with kelp and predators, and undesirable states with sea urchin and calcareous algal dominated kelp barrens, and kelp alone \cite{Steneck_etal_2002}.  The transition from a state with kelp to a sea urchin barren is driven by overgrazing, which can be precipitated by a decline in predators of sea urchins and by climate change \cite{Steneck_etal_2002,Ling_etal_2015,Ling_Keane_2024}. The presence of 'incipient barrens', which is essentially a spatial early warning sign of larger areas without kelp, provides a clear indication of likely tipping.  For kelp systems, the potentially positive news is that management approaches which essentially consist of changing 'parameters' and the state by helping to support predation and removing sea urchins has been shown to help restore kelp forests both experimentally \cite{Christie_etal_2024}, and in an analysis focused on optimizing the approach \cite{ArroyoEsquivel_etal_2023}.

Similar ideas show up in the understanding of coral-algal-grazer systems with the system maintained in the desirable coral dominated state by grazing with the  possibility of tipping if there is not enough grazing which can result from declines in the population of grazing species \cite{McManus_Polsenberg_2004,Mumby_etal_2007}.  As in the kelp example, there is good evidence of different final states, both from degradation at single sites, and from spatial evidence \cite{Mumby_Steneck_Hastings_2013}.  More recent models \cite{Hock_etal_2024} have focused on how stochasticity interacts with the potential for tipping and transients in the neighborhood of the tipping point to lead to an understanding of potential management scenarios that could help with recovery and resilience of coral reefs in the face of challenge brought on by a combination of climate change and anthropogenic stressors \cite{Hughes2003,Hughes2010,Hughes2013,Hughes2018,Hughes2019}.

A terrestrial example with similar dynamics that has been extensively studied both empirically and with models is the transition between forest and savanna.  As in both the preceding examples, there are approaches based on dynamics that are local in space \cite{staver2011,Staver2011b} and spatial models \cite{Baudena2015,vanNes2018} with fire, like grazing in the previous examples, playing a key role.  Human activities that either suppress or increase fire, and climate change that can increase fire, thus play an important role in governing the transition or tipping \cite{hoffmann2012ecological}.The ability to generate observed patterns from simple models \cite{vanNes_etal_2014} suggests that the models can indeed provide insights into management approaches as well as in understanding the forces governing current vegetation patterns and the effect that changing climate alone will have.  For both the coral-algal-grazer system and the forest-savanna system, the potential-flux approach has been applied to the model equations \cite{xu2023non,Xu_etal_2025}.

\section*{Concluding remarks and the roadmap}

The theory of tipping points in complex ecological and environmental systems has made significant progress during the last decade\cite{lenton2023}, however many aspects remain poorly understood or overlooked altogether. Perhaps one of the biggest challenges is understanding  tipping in spatially explicit systems. The tipping mechanisms outlined above (cf.~Table \ref{table1}) have been discovered and studied predominantly in nonspatial models. Models with explicit space can have different properties and, overall, much richer dynamics acquiring features that simply do not exist in their nonspatial counterparts. There is some theoretical and empirical evidence indicating that spatial systems tend to be more resilient to an environmental change due to a variety of mechanisms \cite{olin2024}, in particular due to the effect of self-organized pattern formation\cite{petrovskii2004,malchow2008,Moreno2025}. 
{\color{black}The latter can significantly alter tipping scenarios\cite{vanderVoort2025}, sometimes to the extent where tipping in the strict sense may disappear as an abrupt response to a change in environmental factor (parameter) becomes gradual\cite{rietkerk2021evasion,Marick2026turing}. 
There can be novel tipping mechanisms specific for spatially extended ecosystems. For example, in a fragmented habitat, global tipping can occur when the strength of diffusive coupling between local populations exceeds a critical threshold\cite{Althagafi2021}. 
Also, tipping point can be determined by conditions leading to synchronization between the population dynamics at different locations. In case of large-magnitude population oscillations where the population density falls to a low level, synchronization can lead to system's tipping through turning a local population extinction to the global one\cite{Marick2026dispersal}. In a spatially heterogeneous biome, response to a global environmental change can lead to a fragmented tipping: coexisting states with part of the biome in one state, and part in another\cite{Bastiaansen2022fragmented}. A similar phenomenon 
is known as clustered tipping\cite{Loriani2025}. 
Finally, in a spatially explicit nonautonomous system that accounts for habitat movement (e.g.~northwards, to escape the effect of global warming\cite{Davies2001range,Chen2011rapid}), R-tipping - species extinction - occurs when the movement rate exceeds a critical value\cite{Barker2026arxiv}. These findings provide a quantitative framework for assessing the vulnerability of populations to rapid habitat shifts\cite{Devictor2008birds}. 
}

{\color{black}
Note that, while some aspects of ecosystem complexity (e.g.~broad range of relevant spatial and temporal scales) can also be seen 
in other natural systems, there are system's properties that arguably are special for ecosystems. Perhaps the most obvious ones are the existence of complex foodwebs and changes of species biological traits through adaptation, phenotypic plasticity or evolution. These, respectively, result in tipping mechanisms that are specific for ecosystems and hardly can be seen in abiotic systems: 
(i) disruption of foodwebs can lead to fast and significant biodiversity loss on a regional\cite{Reji2025} and even global scale\cite{Roopnarine2006}, in the latter case potentially leading to mass extinctions; 
(ii) a change in biological traits as a response to environmental change can lead to population's tipping: a fast significant decrease in the population size\cite{Fellows2026bioRrxiv}. 
A mismatch between the rate of environmental change and the rate of species adaptive response leads to ecosystem tipping 
potentially resulting in mass extinctions\cite{Rothman2026}. 
 }

{\color{black}
Note that recent progress in the tipping points science led to a somewhat broader definition of what a tipping point is and/or how system's tipping actually occurs. 
Recall that, according to the original understanding of tipping phenomena\cite{lenton2008tipping,ashwin2012tipping}, in the absence of noise the existence of a critical threshold (in a bifurcation parameter for B-tipping or in the rate of change for R-tipping or in the magnitude of the shock for S-tipping) is a necessary condition for the system to tip. However, it becomes somewhat different if the system state before or after tipping is more complicated than a steady state, e.g.~periodic (limit cycle) or chaotic (strange attractor). In that case, there is no single threshold as system's response to perturbations may depend on its position on the attractor. Instead, tipping is described by a distribution of critical thresholds, which is captured by the recently developed concept of partial tipping\cite{Alkhayuon2018,Kaszas2019,Basak2024partial,Lohmann2021}.
 }

{\color{black}
In the presence of noise, a critical threshold is not required. 
Interestingly, in the updated list of tipping mechanisms (cf.~Table \ref{table1}), a threshold is not a necessary condition for tipping to occur even in the absence of noise. 
In particular, in case tipping occurs at the end of long transient dynamics (LT-tipping)  
due to a ghost attractor\cite{hastings2018transient}, the existence of multiple scales (e.g.~through slow-fast dynamics\cite{morozov2020}) or the effect of saddle-points\cite{Ashwin2005} (including chaotic saddles\cite{Grebogi1983,Grebogi1986,do2004,Duarte2012}) a threshold is not needed.}


The immense complexity of their structure and dynamics makes ecosystems different from the majority of physical systems and poses a significant challenge for  tipping point theory. 
{\color{black}Moreover, ecosystems do not exist in isolation. 
All major systems on Earth are coupled and that makes their  complexity even higher.
This applies not only to the climate-biosphere system, where coupling between biotic and abiotic components is widely accepted\cite{lenton2008tipping,Charlson1987,Kleidon2010}, but also to somewhat less appreciated, more general, coupled socio-economical-ecological system\cite{Liu2007,Farahbakhsh2022}. 
There is growing evidence of a feedback} that `positive' regime shifts in social and/or economic systems can provide for the global climate and environment\cite{Lenton2013,lenton2020}.
In particular, a change in human perception of the ongoing climate/ecological crisis and, respectively, a change in  human behaviour (including their attitude to consumption of goods, diet, and change in  life style more generally) was shown to have a sizeable capacity to slow down  global environmental change\cite{Beckage2018}. 

Thus, despite the significant progress made over the last decade in understanding and predicting tipping points, {\color{black}in particular in ecosystems, many open questions remain and new challenges are emerging. 
Firstly, unlike abiotic systems, ecological systems exhibits remarkably high variability at all levels of organization. 
For example, individuals of the same species are not identical and hence can have different tolerance to a change in their ambient environment. Intuitively, due to such variability and adaptivity, the system response to a change can be gradual rather than abrupt and hence may avoid tipping. However, good understanding of that is lacking. }

Secondly, many models of biological and ecological systems are known to show structural sensitivity to parametrization of interspecific interactions, {\color{black} where implementing two different functions with very close shapes in a model would result in distinctly different model predictions, which is observed across wide range of parameters \cite{Fussmann2005,Adamson2013,Manna2026}. How the mentioned property of structural sensitivity} can affect models' capacity to describe tipping is an entirely open question. 

Thirdly,
{\color{black}in this review paper our main focus is on theory and corresponding mathematical models. Indeed, majority of existing studies on tipping points in ecosystems are model-based rather than data-driven.
One reason for that is that data of required length and quality are rarely available. Moreover, ecosystems are usually affected by strong noise, which can make the tipping points difficult to detect and even more difficult to anticipate\cite{Hastings2010regime}. 
Having performed an extensive meta-analysis of empirical data, 
a recent study\cite{Hillebrand2020} showed that threshold-type ecosystem responses to a minor variation in forcing were rarely detectable. The observed responses were more typically characterized by gradual changes rather than abrupt ones. 
Qualitatively, this may be explained by the prevalence of partial typing\cite{Alkhayuon2018,Kaszas2019,Basak2024partial,Lohmann2021}
in real-world systems (which, given their great variability, can be intuitively expected). However, a quantitative investigation into this issue is entirely missing and this poses another big challenge for future research.}

We also mention here that the situation with ecological data has changed significantly over the last decade as ``big data'' for many  ecosystems have become available\cite{Hampton2013,Wagele2022,Abrams2025}.
Although the data quality still remains an issue, that opens a possibility for a broader use of data-driven methods, which can in principle provide a powerful alternative to model-based approaches\cite{Bury2021,Zhuge2025}. This is particularly true for studies on ecosystem dynamics where it is hardly possible to include all the ecological complexity in the model, and hence the model construction can be subjective. 


Understanding of tipping mechanisms and scenarios and the corresponding early warning signals that accommodate the above challenges should be the overall direction of future research.

\section*{Acknowledgments}
VL acknowledges the partial support provided by the Horizon Europe Projects Past2Future (Grant No. 101184070) and ClimTIP (Grant No. 100018693), by the ARIA SCOP-PR01-P003—Advancing Tipping Point Early Warning AdvanTip project. SP and VL are also supported by the European Space Agency Project PREDICT (Contract 4000146344/24/I-LR). AH is supported by US NSF Grant 2025235.


\section*{Author contributions}
The authors contributed equally to all aspects of the article. 

\section*{Competing interests}
The authors declare no competing interests. 

\section*{Publisher’s note}
Springer Nature remains neutral with regard to jurisdictional claims in published maps and institutional affiliations.

\end{document}